\documentstyle{mn}
\title[MM/Submm Polarisation of Compact Radio Sources] {A Millimetre/Submillimetre Polarisation Survey of Compact Flat-Spectrum Radio Sources.}

\author[R. Nartallo {\em et al.}] {
  R. Nartallo$^{1,3}$, W. K. Gear$^2$, A. G. Murray$^3$, E. I Robson$^4$ and 
     J. H. Hough$^5$ \\
  $^1$ University of Edinburgh, Blackford Hill, Edinburgh EH9 3HJ, U.K. \\
  $^2$ Royal Observatory, Blackford Hill, Edinburgh EH9 3HJ, U.K. \\
  $^3$ Department of Physics, Queen Mary and Westfield College, Mile End Road, 
       London E1 4NS, U.K. \\
  $^4$ Joint Astronomy Centre, 660 North A'oh\={o}k\={u} Place, University Park,
       Hilo, Hawaii 96720, USA \\
  $^5$ Division of Physical Sciences, University of Hertfordshire, 
       Hatfield, U.K. \\
}
\date{
  Accepted; $\;\;\;\;$  
  Received; $\;\;\;\;$  
  in original form 
}

\begin{document}

\maketitle

\begin{abstract}
We present the results of the first survey and monitoring study of the linear polarisation properties of compact, flat-spectrum radio sources at 
millimetre/submillimetre wavelengths and discuss the implications of the inferred magnetic field structure for the emission models involving shock waves in relativistic jets on sub-parsec scales.  We find significant polarisation in most sources but, in general, the magnetic field on sub-parsec scales is less well ordered than on parsec scales.  We observe no difference in polarisation properties between the BL Lac objects and compact flat-spectrum quasars at these wavelengths.  Although we find the behaviour of some sources, particularly the most highly polarised, to be very consistent with the predictions of transverse {\em shock-in-jet} models, the detailed behaviour of most objects is not.  Conical shock structures can more readily explain the observed diverse behaviour of the sample, although some degree of bending of the jet may still be necessary in some cases.

\end{abstract}

\begin{keywords}
BL Lacertae objects : polarisation --- quasars : polarisation  ---  galaxies : jets : shocks --- radio continuum : galaxies --- submillimetre : JCMT : UKT14 
\end{keywords}

\section{Introduction}
Compact, flat-spectrum, radio-loud extragalactic sources, sometimes also known as `blazars' (Angel \& Stockman 1980, Antonucci 1993), exhibit many common characteristics.  In particular, they show core-jet morphology in VLBI images with a succession of synchrotron self-absorbed knot-like structures combining to give the overall flatness of the radio spectrum, and frequently exhibit apparent superluminal motion along with large-amplitude variability at all wavelengths from radio to gamma-rays.  Significant and variable linear polarisation is also seen at all wavelengths where measurements have been attempted.  At millimetre/submillimetre wavelengths the emission arises through the synchrotron mechanism in the core regions of the jets, i.e. close to their origin (see e.g. Gear 1988), which are completely unresolved on even the highest-frequency VLBI maps.  Photometric monitoring in this band has greatly improved our understanding of the physical processes in the sub-parsec scale jets and has led to the development of the canonical {\em shock-in-jet} paradigm (Marscher \& Gear 1985 and Hughes, Aller \& Aller 1985, 1989a, 1989b).  An essential ingredient of jet models that has so far been missing is the geometry of the magnetic field on these very small scales and how it may vary on different physical scales.  The millimetre/submillimetre emission is expected to be linearly polarised if the magnetic field is ordered and, with very rare exception (e.g. Robson {\em et al.} 1983), it is always optically thin.  This means that internal synchrotron self-asorption and external Faraday rotation of the polarisation vector can be ignored.  Therefore, we can reliably assume that the magnetic field direction must lie perpendicular to the observed polarisation position angle (see e.g. Pacholczyk 1970). 

The advent of polarisation-sensitive VLBI imaging has been a powerful tool in investigating magnetic field behaviour (e.g. Cawthorne  {\em et al.} 1993, Gabuzda  {\em et al.} 1994).  However, no information has been available on the degree of order and the orientation of the magnetic field close to the point of creation of the jet.  In the early 1990s we began a 1.1 and 0.8 mm polarisation survey of a significant sample of flat-spectrum radio sources, most of which have subsequently been measured over several epochs at 1.1 mm.  In this paper we present the data from this survey, which allow us to carry out the first investigation of the magnetic field properties in the core of compact radio jets, and we discuss the implications for models of these sources.

\section{Sample}

Polarisation detections have been obtained on 26 different sources, half of which have been observed at 4 or more different epochs between 1991 and 1996 and single-epoch upper limits have been measured on a further two objects.  A complete list of the objects with their polarimetric observations to date is given in Table~1.  The main criterion adopted for the selection of sources has been that their flux density at 1.1 mm should be at least 1 Jy so that a detection can be achieved in a reasonable time.  Due to the highly variable nature of these objects and in order to ensure that changes in the emission between epochs are detected above the uncertainty level of the measurements, a $5\sigma$ polarimetric detection was always attempted and often achieved.  Other selection criteria include the availability of published high-frequency VLBI observations, access to the target sources at various times of the year and keeping a balanced mixture of different optical types.

\section{Observations}

\subsection{Data Acquisition}
The observations were acquired using the Aberdeen/QMW Polarimeter (Murray 1991 and Murray  {\em et al.} 1992) in conjunction with the UKT14 bolometric receiver (Duncan  {\em et al.} 1990) located at the east Nasmyth focus of the 15-m James Clerk Maxwell Telescope (JCMT) on Mauna Kea, Hawaii.  The polarimeter consists of a rotating half-waveplate followed by a fixed wire-grid analyser.  It is operated in the `Step and Integrate' mode, whereby a standard photometric measurement is performed at a number of equi-spaced positions of the waveplate (between $0\degr$ and $360\degr$), thus modulating the polarised component in the signal.  The observations are automated by adopting a waveplate rotation and beam switching scheme which drives the waveplate and nods the telescope at each waveplate position.  The telescope's secondary mirror is used to chop between the source and the sky thermal background at the default frequency of 7.8125 Hz with a chop throw of $60\arcsec$.  The position angle of the linear polarisation vector is rotated by the parallactic angle of the source due to the alt-azimuth mounting of the telescope and it is further rotated by its elevation in being reflected off the tertiary mirror.  These rotations can be corrected for either during acquisition or {\em off-line} in software.  When making purely photometric observations with the polarimeter placed in front of the receiver, two separate measurements with the waveplate at $0\degr$ and $45\degr$ are taken and averaged.  Further technical details are given in Murray (1991) and Nartallo (1995).

\subsection{Data Reduction}

\begin{figure}
 \vspace{13.5cm}
 \includegraphics{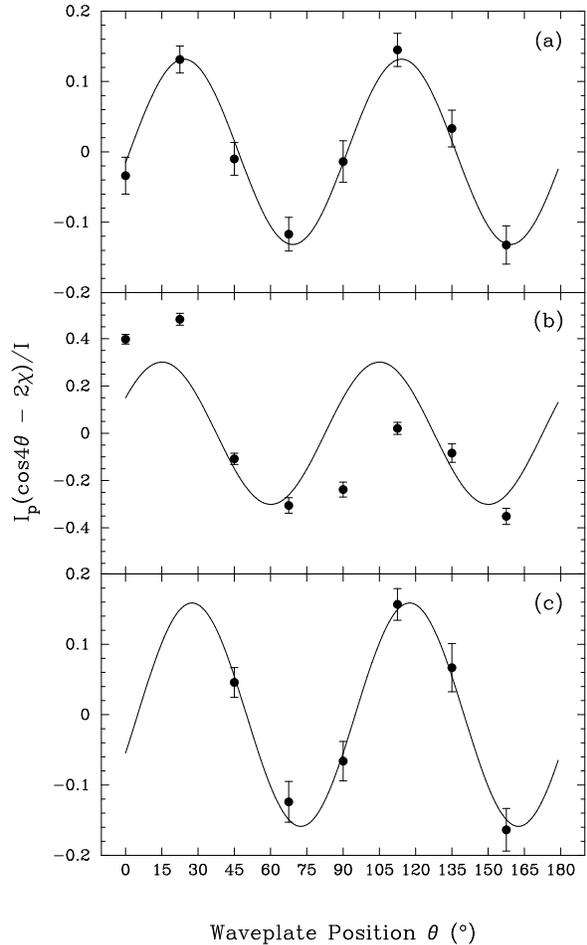}
 \caption{Examples of the fitting and noise removal procedures.  Plot (a) shows a good fit to a
          noise-less first half-cycle of the waveplate, while plots (b) and (c) show fits
          to the second half-cycle of the same observation before and after applying the noise removal algorithm.}
 \label{fits}
\end{figure}

The raw output of the receiver $S$ can be represented as a function of the waveplate position $\theta$ by the following equation (e.g. Nartallo 1995),

\begin{equation}
 S(\theta) = \frac{1}{2} \, [ I_{u} + I_{p} [ 1 + \cos(4\theta - 2\chi) ] ] 
\end{equation}

\noindent 
where $I_{u}$ and $I_{p}$ are the intensities of the unpolarised and polarised components of the signal respectively, and $\chi$ is the position angle of the linear polarisation vector.  In terms of the Stokes parameters $I$, $Q$ and $U$ (which fully describe linear polarisation) this can be written as:

\begin{equation}
 S(\theta) = \frac{1}{2} \, [ I + Q\cos(4\theta) + U\sin(4\theta) ] 
\end{equation}

\noindent 
where $I=I_{u}+I_{p}$, $Q=I_{p}\cos(2\chi)$ and $U=I_{p}\sin(2\chi)$.  The Stokes parameters are estimated by performing a least squares fit on a number of data points taken at different positions of the waveplate.  This technique allows the removal of individual spikes which are occasionally observed in the raw data due to changing sky emissivity, cosmic rays or electronics glitches.  A noise removal algorithm was developed that optimises the $\chi^{2}$ value of an $n$-point fit {\em while maximizing} the number of points used within a waveplate cycle.  Other methods tried for determining the Stokes parameters include fitting the data by a Fast Fourier Transform and the estimation of $Q$ and $U$ directly from photometric measurements taken $45\degr$ apart.  However, spiked points cannot be taken out when using these techniques, often resulting in unusable estimates of $Q$ and $U$.

In practice, a typical observation consists of 16 waveplate positions that are fitted in half-cycles of 8 points.  Half-cycles that have more than 3 points removed are thrown away.  Figure~1 illustrates a fit to a 16-point observation of a highly polarised source ($\sim 15\%$ with position angle $40\degr$).  Plot (a) shows the normalised raw data for the first half-cycle and its corresponding fit and plot (b) is the equivalent representation for the second half-cycle, which is clearly affected by noise.  The peak of the fitted curve gives the percentage polarisation and its displacement from the $\theta = 0\degr$ position is half the position angle of the polarisation vector.  While the fit shown in plot (a) gives a degree of polarisation of $13\%$ and a position angle of $45\degr$, the values obtained from the second half-cycle shown in plot (b) are $31\%$ and $61\degr$ respectively.  Plot (c) shows the result of applying the noise removal algorithm to (b) and it is now seen that it is in good agreement with (a) (giving $15\%$ polarisation and $38\degr$).

After successfully fitting the raw data the normalised Stokes parameters $q$ and $u$ are calculated and corrected for the Instrumental Polarisation (IP) and rotations.  The main contributor to the IP is the protective wind blind of the JCMT (Flett \& Murray 1991) and it is therefore dependent on the actual blind used and its tension at each epoch (further details of the origin and characteristics of the IP are discussed in Murray 1991).  The value of the IP to be subtracted from observations of point sources such as blazars is estimated from observations of planets with small angular size (e.g. Uranus, Mars or Neptune) which are considered to be unpolarised at millimetre wavelengths.  Typically, the IP measured on these planets has values in the range $1.5 - 1.8\%$ at 1.1 mm and close to $0.5\%$ at 0.8 mm, which remain largely unchanged over the period of an observing run.  Unlike the IP position angle (which tracks with elevation) the level of linear instrumental polarisation does not have an elevation dependance.  The IP position angle at zero elevation is $\sim 173\degr$.

Independent estimates of $q$ and $u$ are obtained from several revolutions of the waveplate and checked for homogeneity by performing a Kolmogorov-Smirnov test, that compares their distributions with a gaussian distribution and can indicate the presence of spurious observations.  All the observations that are finally accepted are averaged to improve the statistical determination of the polarisation parameters.  The number of observations available for each source and their consistency allow an error weighted average to be used in most cases, which improves the final signal to noise.  If only a few measurements are available and/or their individual signal to noise varies significantly between them, a straight mean is used instead. 

Finally, the degree of linear polarisation $P$ and its uncertainty $\sigma_{p}$ can be obtained from the averaged normalised Stokes parameters using the expressions,

\begin{equation}
 P_{obs} = \sqrt{q^{2} + u^{2}}
\end{equation}

\begin{equation}
 \sigma_{p} = \frac{\sqrt{ (q\times\sigma_{q})^{2} + (u\times\sigma_{u})^{2} }}
              {P_{obs}}
\end{equation}

\noindent
the observed degree of polarisation $P_{obs}$ is then corrected for bias using the Wardle and Kronberg method (Wardle \& Kronberg 1974) which is adequate for $P_{obs}/\sigma_{p} \geq 0.7$ (Simmons \& Stewart 1985),

\begin{equation}
 P =  \sqrt{ P_{obs}^{2} - \sigma_{p}^{2} }
\end{equation}

\noindent
The position angle $\chi$ and its uncertainty $\sigma_{\chi}$ are given by,

\begin{equation}
 \chi = \frac{1}{2} \, \arctan \left ( \frac{u}{q} \right )
\end{equation}

\begin{equation}
 \sigma_{\chi} =  \frac{1}{2} \, \frac{\sigma_{p}}{P}
\end{equation}

\noindent
where the expression for $\sigma_{\chi}$ is a good approximation if $P \gg \sigma_{p}$ (Naghizadeh-Khouei \& Clarke 1993).  (A more detailed account of all the data reduction algorithms used and tests carried out on them is given in Nartallo 1995).

\subsection{Observing Wavelengths}

\begin{figure*}
 \vspace{13.25cm}
 \includegraphics{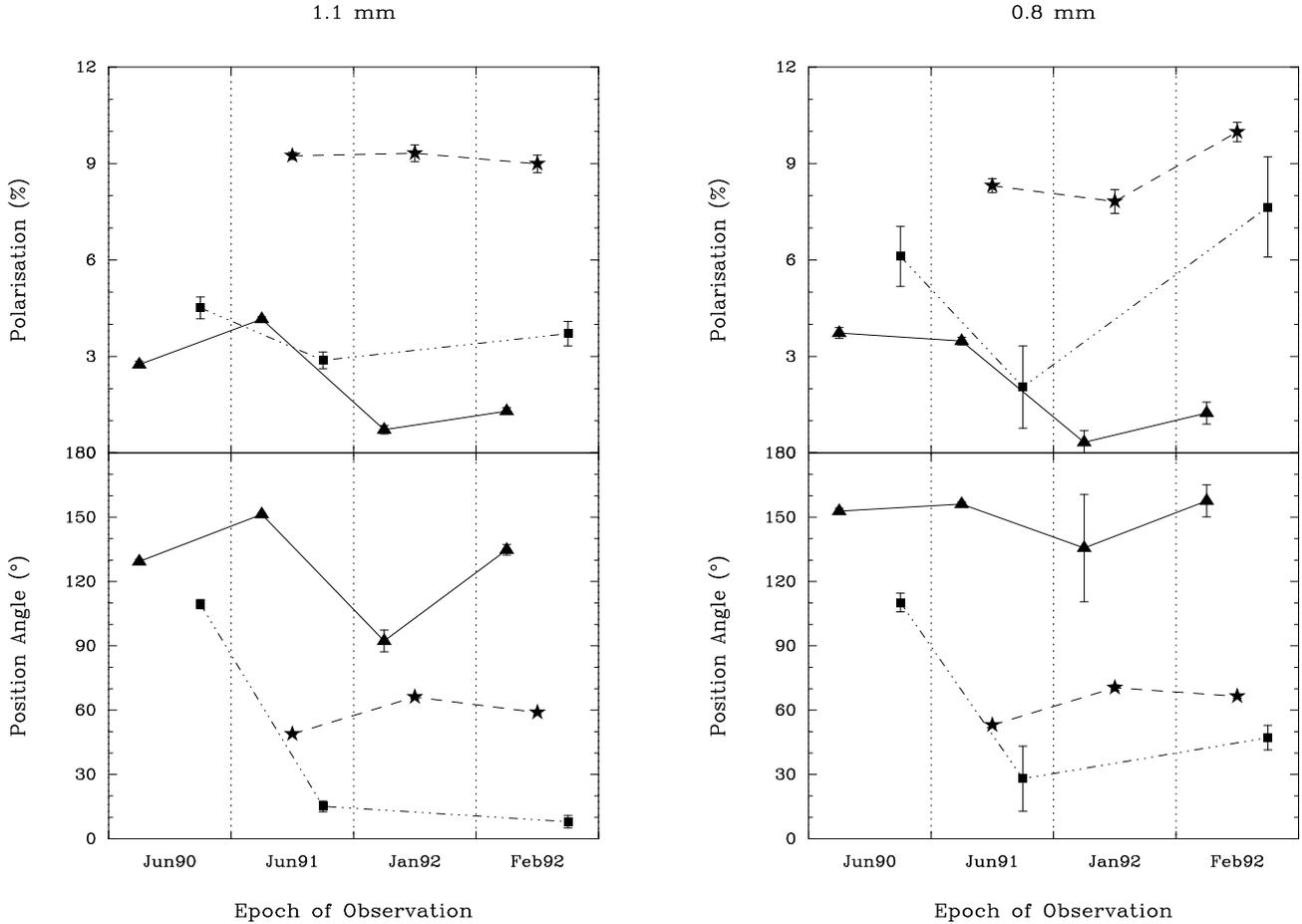}
 \caption{Comparison between the 1.1 and 0.8 mm polarisation properties of three compact, flat-spectrum radio sources.  The data represented by 
          triangles joined by a solid line correspond to observations of 3C273, the stars and dashed lines correspond to 3C279 and the squares and 
          dashed-dotted lines are measurements of 3C454.3.  The error bars shown represent the uncertainty in each individual detection, as given in
          Table~1.}
 \label{wcomp}
\end{figure*}

Polarimetric observations at 1.1 and 0.8 mm have been obtained at nine and three different epochs respectively.  A few extra observations of selected sources taken during the two commissioning runs of the polarimeter (Flett \& Murray 1991) have been added to the database and re-analysed.  Initially, observations were systematically made at the two wavelengths, thus limitting the number of objects that could be covered in a single epoch.  Data from these earlier epochs show that the polarisation properties of most sources at 1.1 and 0.8 mm are very similar.  This is illustrated by Figure~2, where the variability with epoch of the percentage linear polarisation and position angle is shown for three sources at these two wavelengths.  The plots include four-epoch observations of 3C273 and three-epoch observations of 3C279 and 3C454.3.  These three objects have different overall polarisation properties, ranging from lowly polarised (3C273) to very highly polarised (3C279).  Since there is little knowledge to gain by monitoring the polarisation at the two wavelengths, it was decided to concentrate on single-frequency polarimetry at 1.1 mm in order to make the sample statistically significant by increasing the number of sources measurable during an observing run.  The choice of wavelength was determined by the fact that a good signal to noise can be achieved more quickly at 1.1 mm and also this atmospheric window is less affected by weather conditions.  However, the level of instrumental polarisation is a factor 3 greater than at 0.8 mm.

Whenever possible, photometric measurements of the sources have also been taken at 2.0, 1.1 and 0.8 mm, from which a three-point spectral index is calculated (the source spectrum $I(\nu) \propto \nu^{\alpha}$).  For the earlier epochs, the photometry has been extracted from the 1.1 and 0.8 mm polarisation observations and the spectral index is therefore estimated from two wavelengths only (thus having greater uncertainties).  The photometric measurements are flux calibrated against planets measured close in time and at similar airmasses (Griffin {\em et al.} 1986, Orton {\em et al.} 1986, Griffin \& Orton 1993).

\section{Results}

The polarisation and photometric measurements obtained on each source during the course of this monitoring programme are tabulated in Table~1 for each observation epoch.  The table includes the percentage linear polarisation $P$ and position angle $\chi$ at 1.1 and 0.8 mm, the flux density $F$ in Janskys at 2.0, 1.1 and 0.8 mm and the spectral index $\alpha$ determined from the available wavelengths.  In Figure~3 we have plotted the four measured parameters against epoch of observation for every source detected.  The polarisation position angle plots are centred about the position angle of the VLBI jet as given in Table~2, indicated by a horizontal dashed-dotted line; the direction perpendicular to the jet is indicated by two dotted lines at $90\degr$.  The inferred magnetic field geometry is perpendicular to the jet if $\chi$ lies close to the VLBI jet direction and parallel if they are closer to the dotted lines.  A summary of the main features observed in the 1.1 mm measurements obtained on each source is given in Appendix~A.

\begin{figure*}
 \begin{center}
 \end{center}
 \vspace{220mm}
 \includegraphics{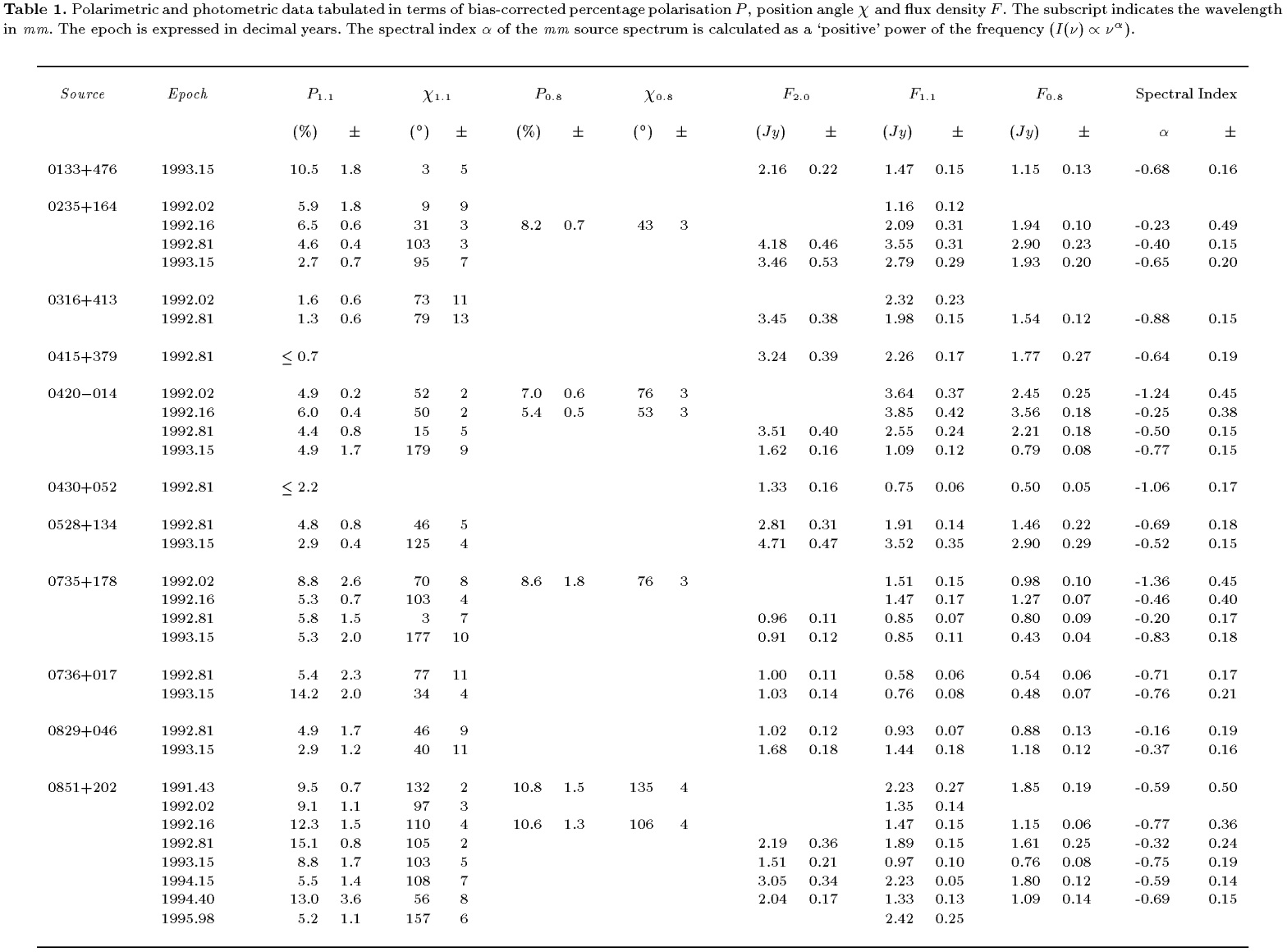}  
\end{figure*}

\begin{figure*}
 \begin{center}
 \end{center}
 \vspace{220mm}
 \includegraphics{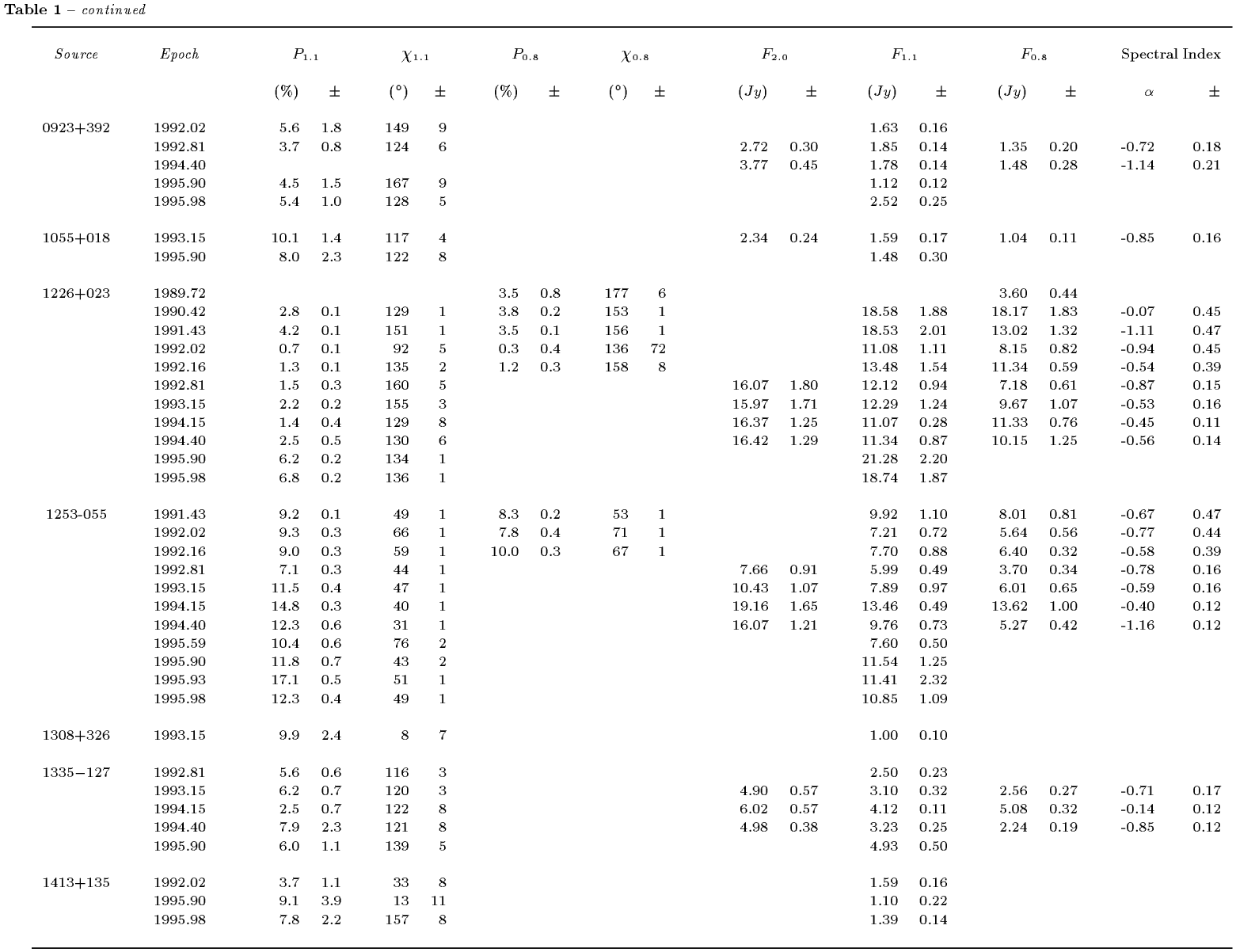}  
\end{figure*}

\begin{figure*}
 \begin{center}
 \end{center}
 \vspace{220mm}
 \includegraphics{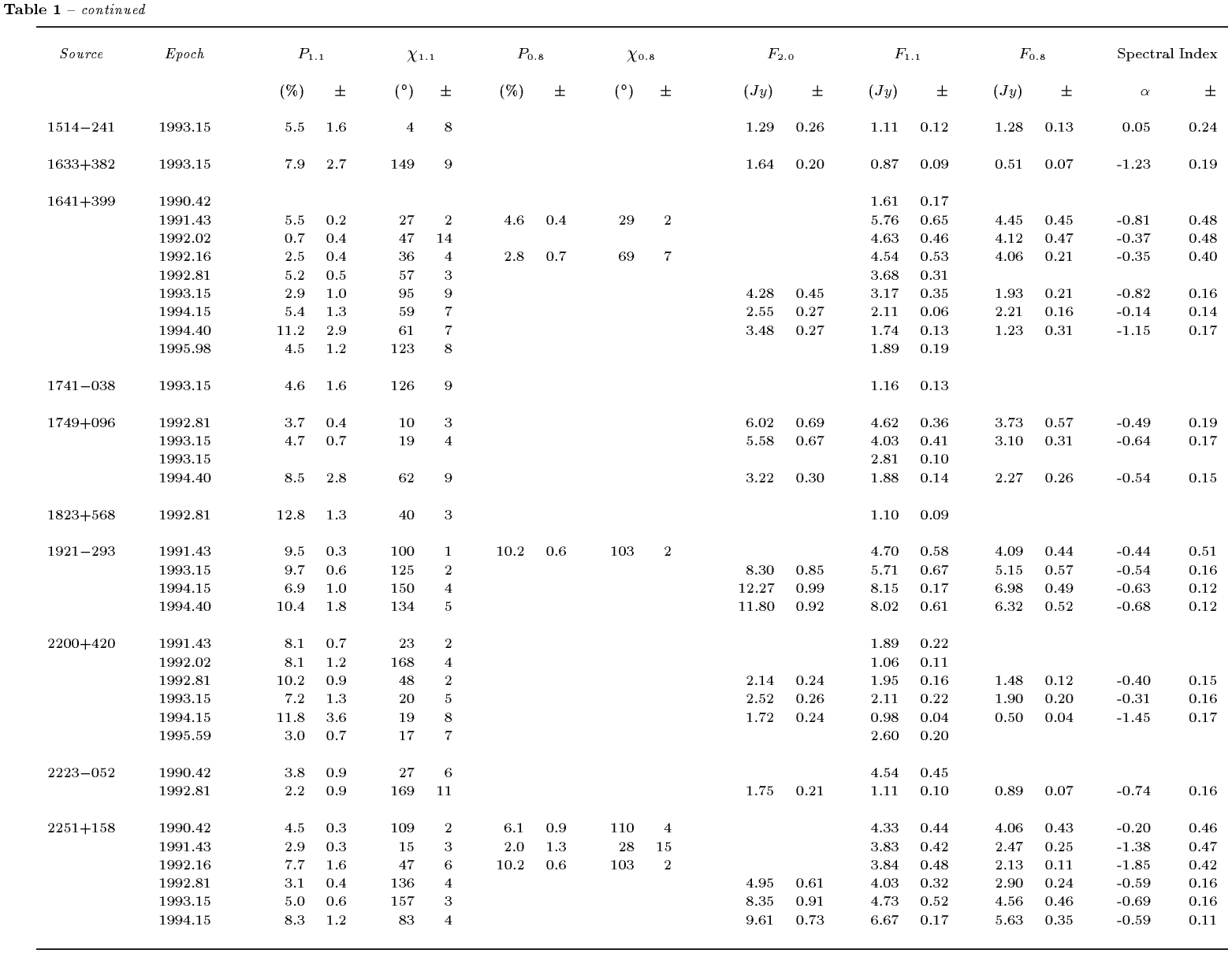}  
\end{figure*}

\begin{figure*}
 \vspace{10.15cm}
 \includegraphics{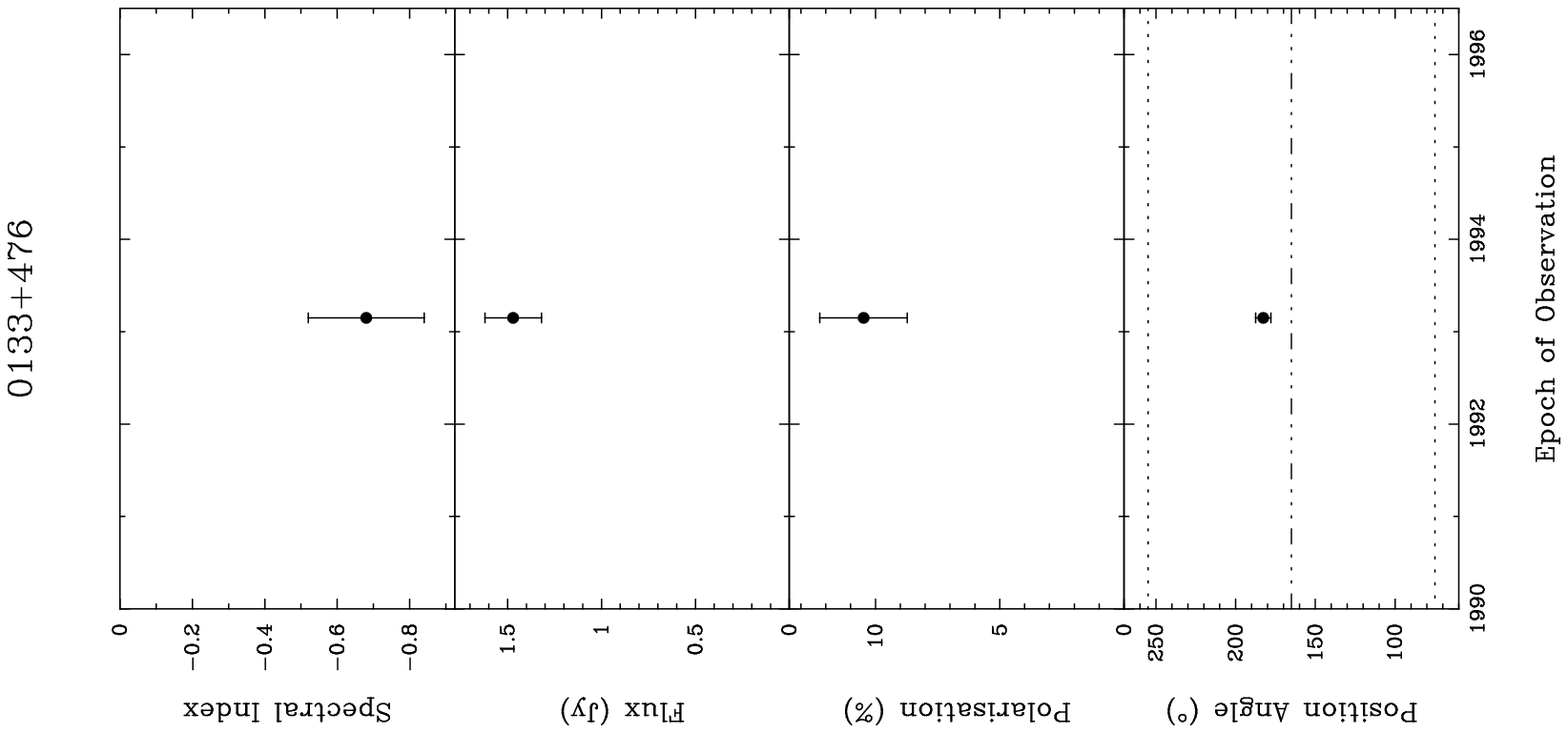} 
 \includegraphics{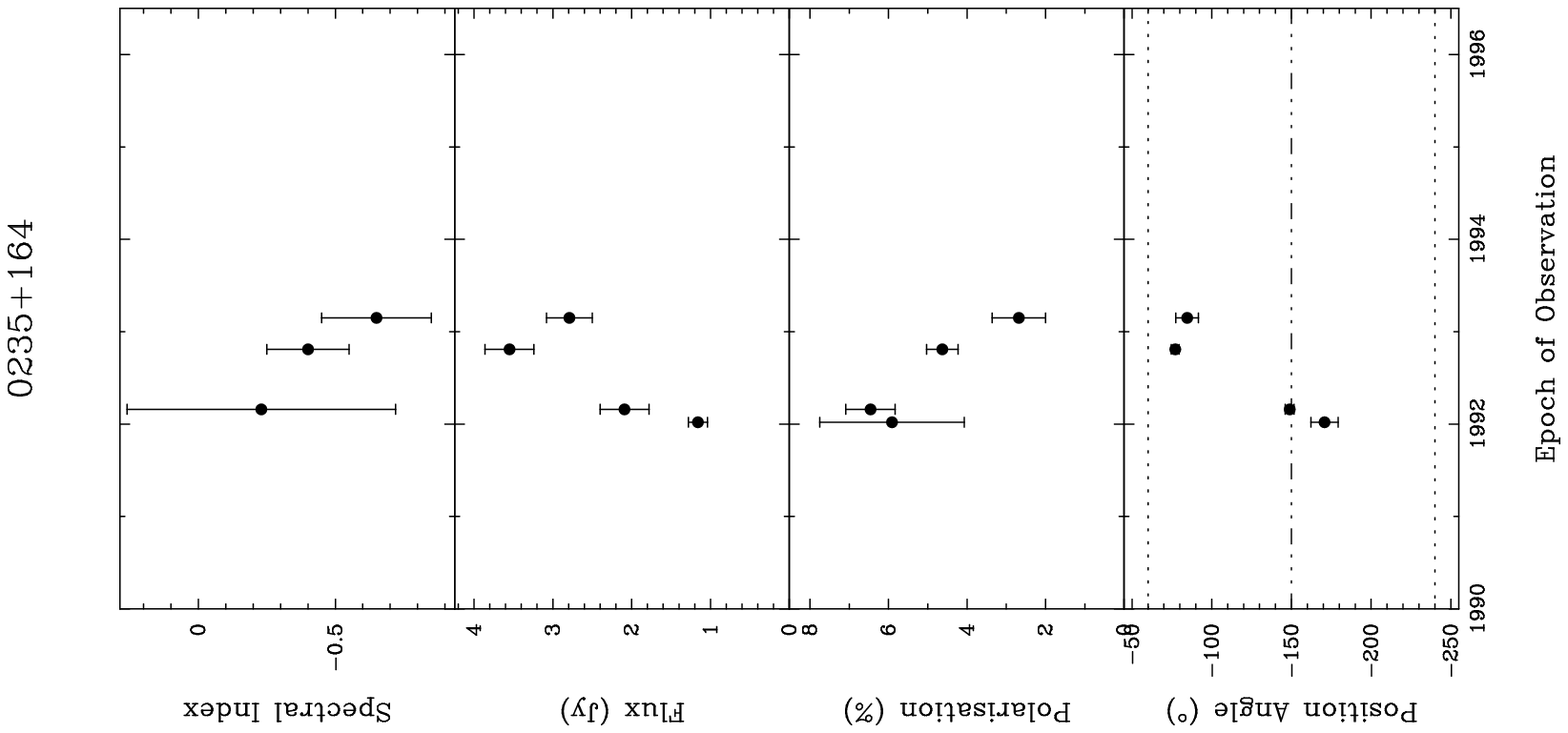} 
 \includegraphics{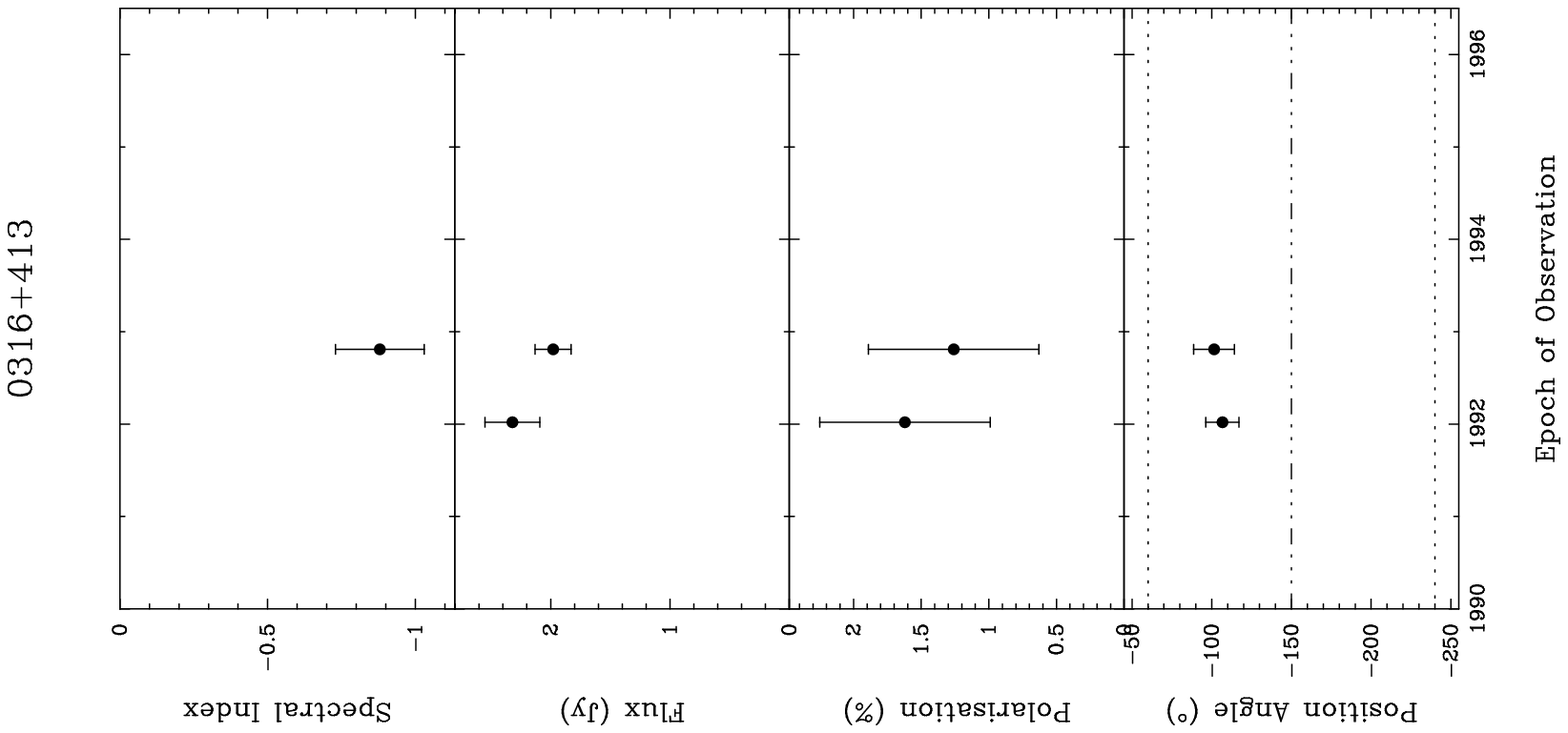} 
 \vspace{11.75cm}
 \includegraphics{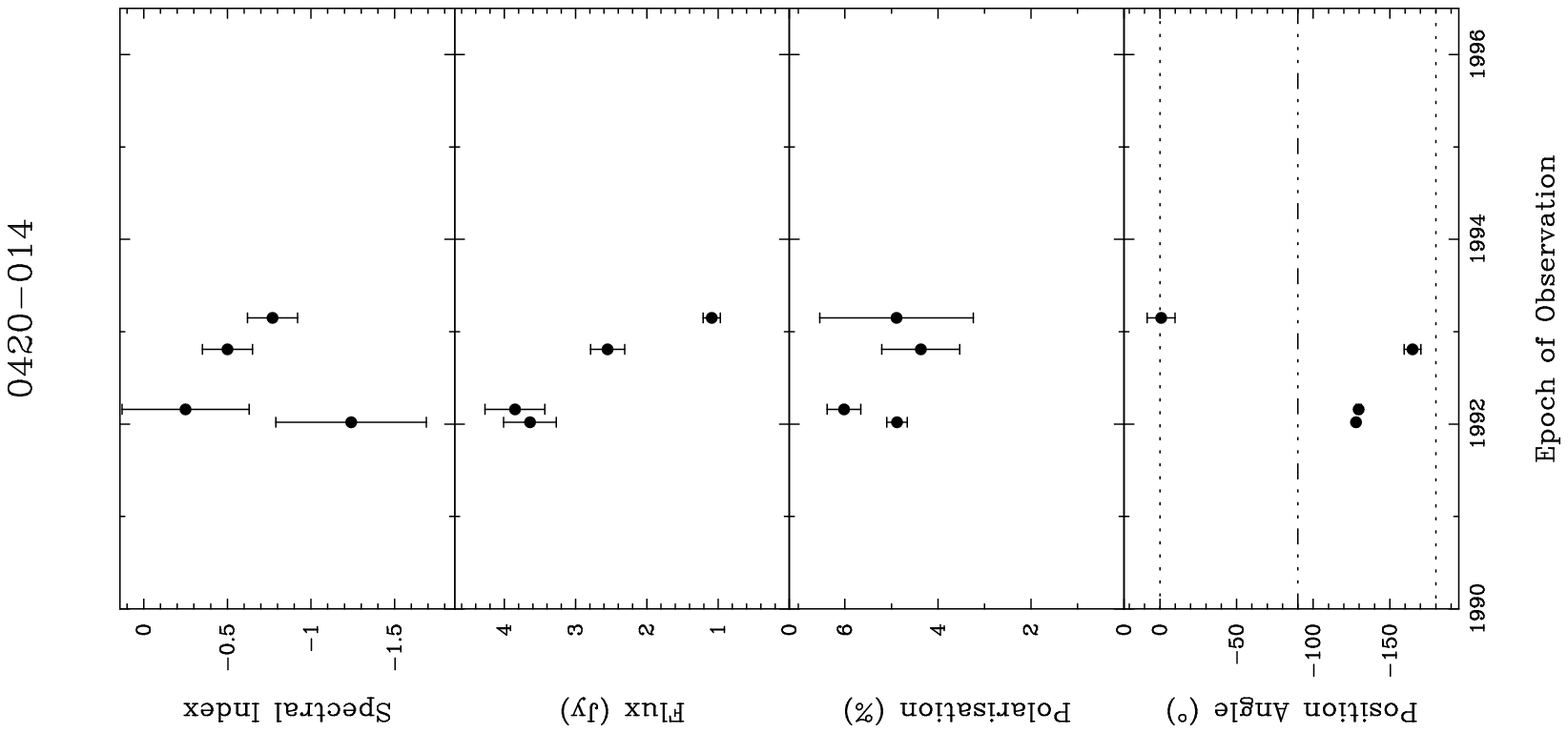} 
 \includegraphics{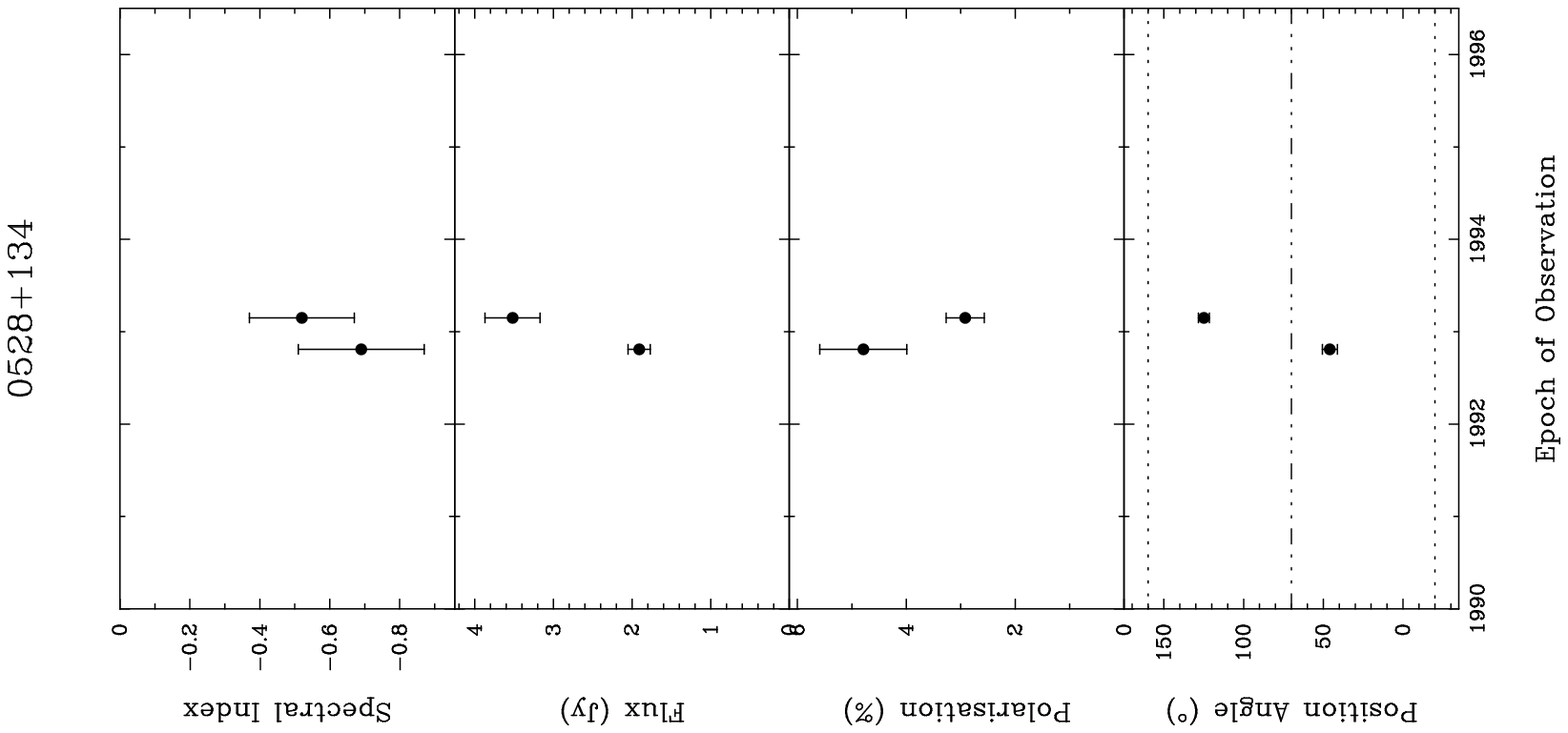} 
 \includegraphics{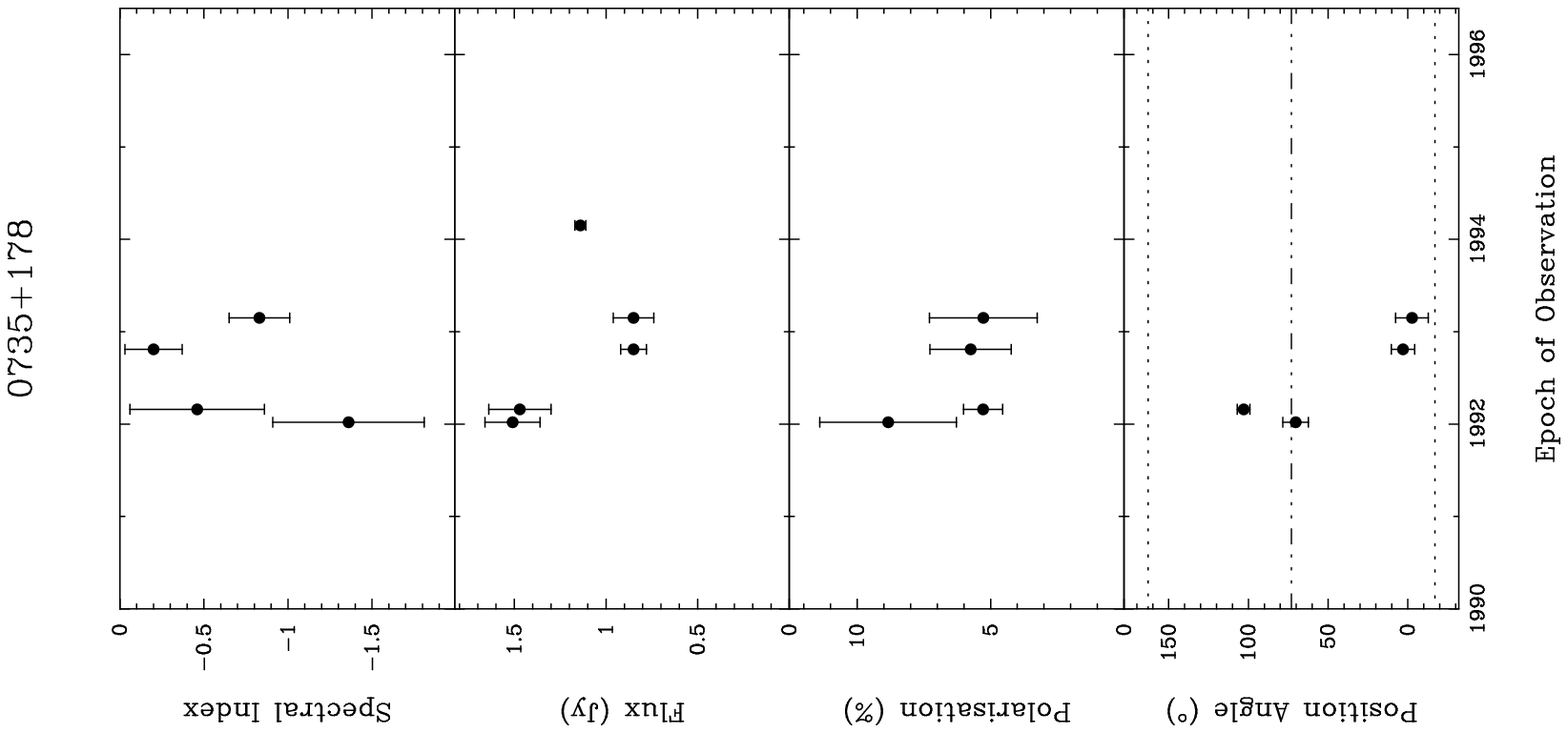} 
 \vspace{1.5cm}
 \caption{The {\em mm} spectral index and the flux, percentage polarisation and 
          position angle measured at {\em 1.1 mm} is plotted against epoch.}
\end{figure*}

\begin{figure*}
 \vspace{10.15cm}
 \includegraphics{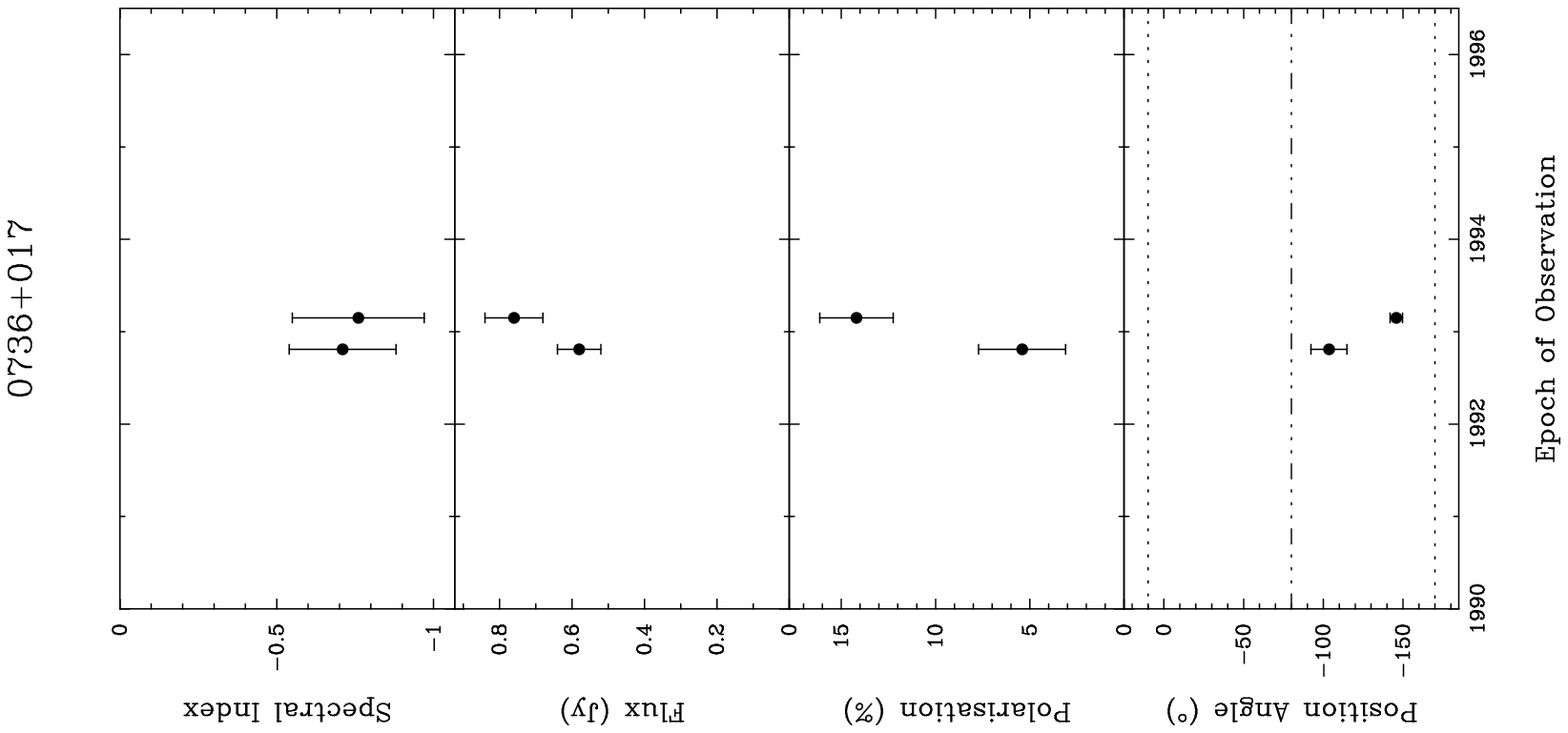} 
 \includegraphics{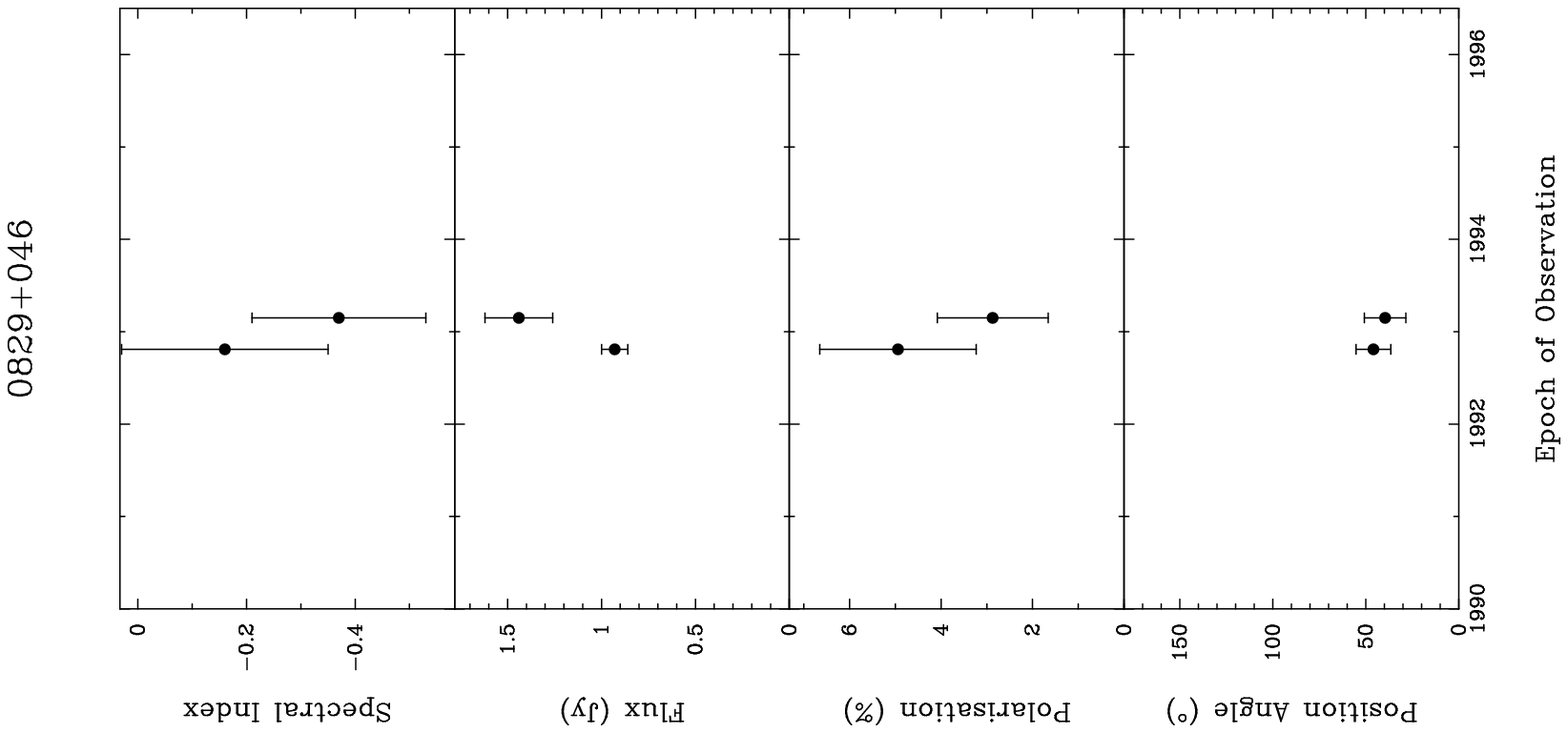} 
 \includegraphics{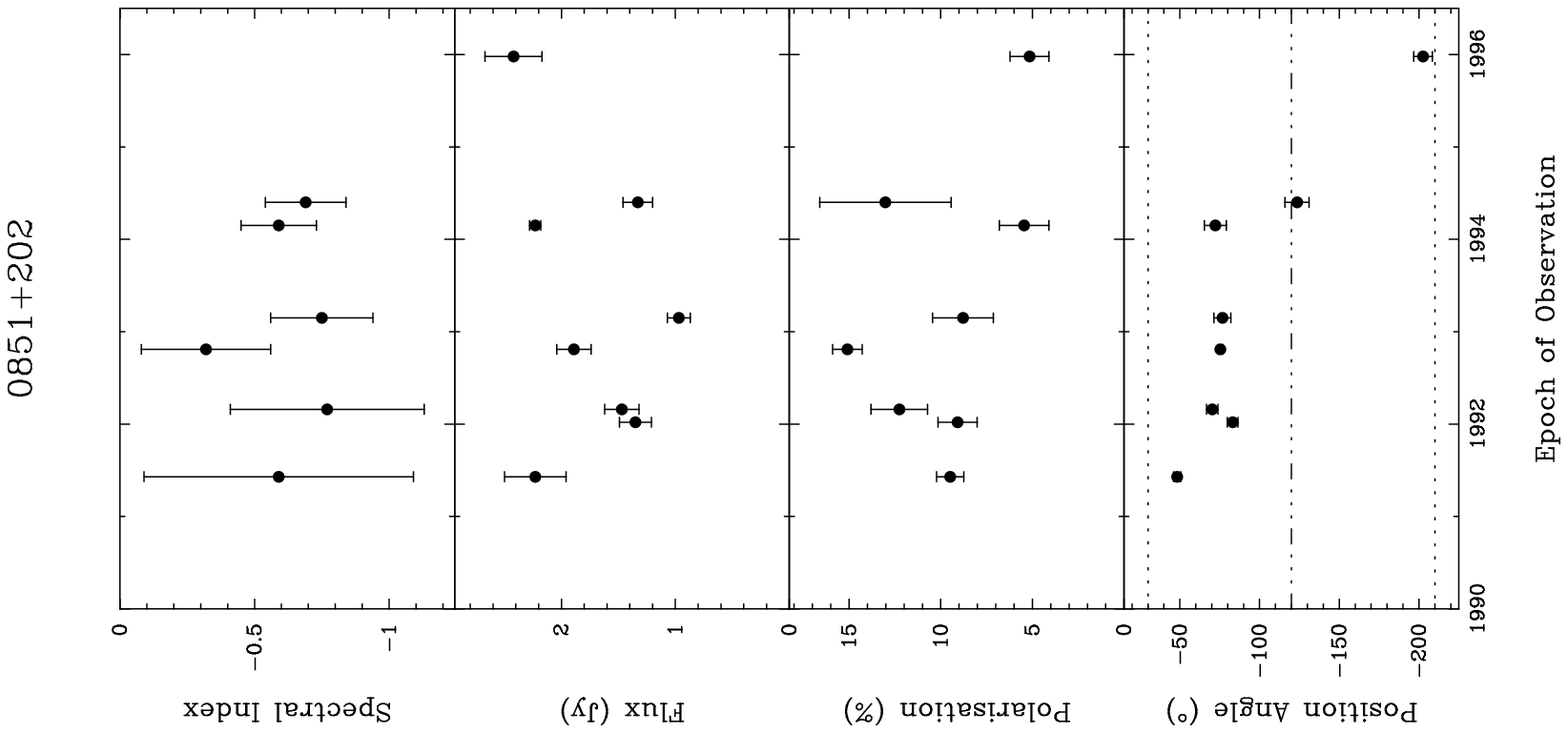} 
 \vspace{11.75cm}
 \includegraphics{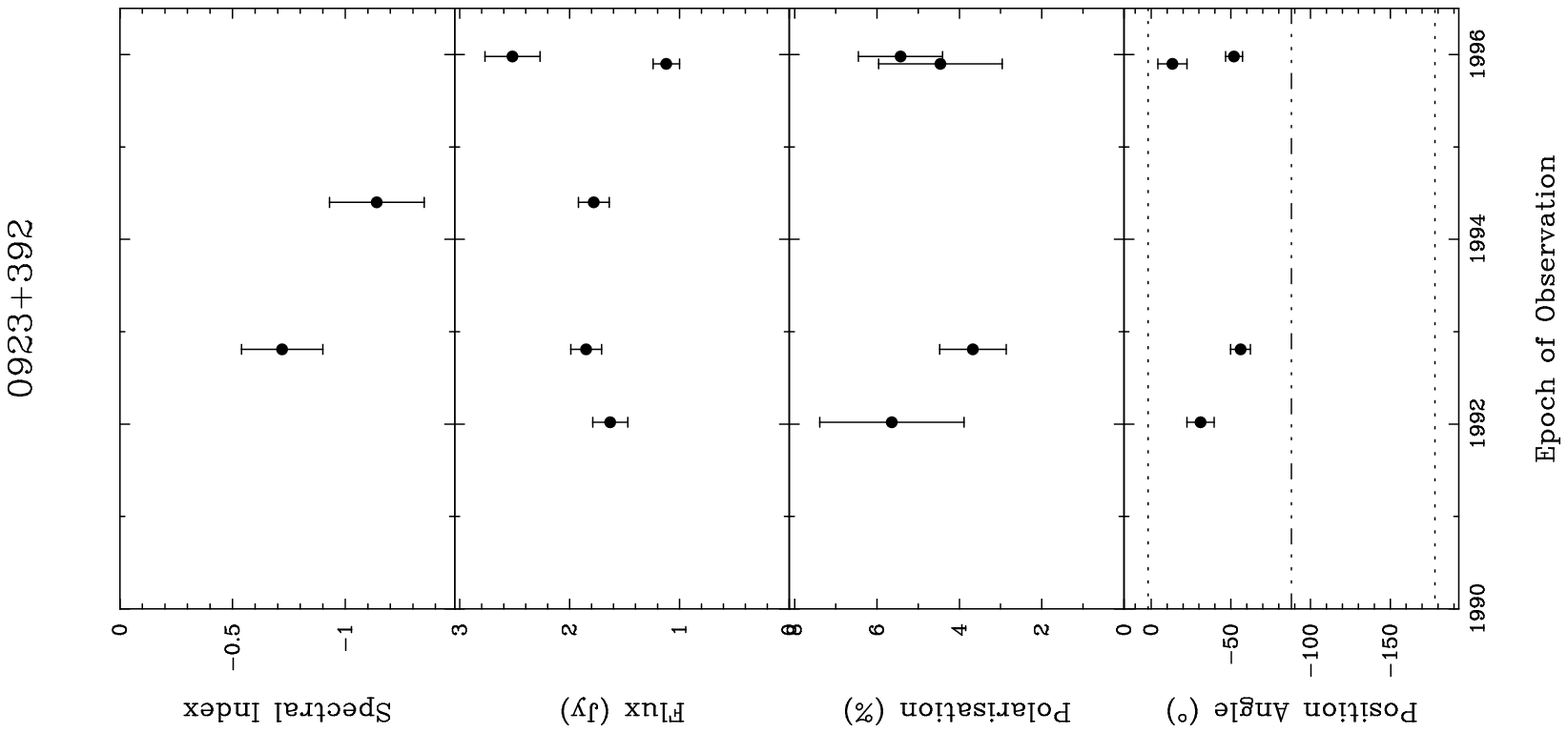} 
 \includegraphics{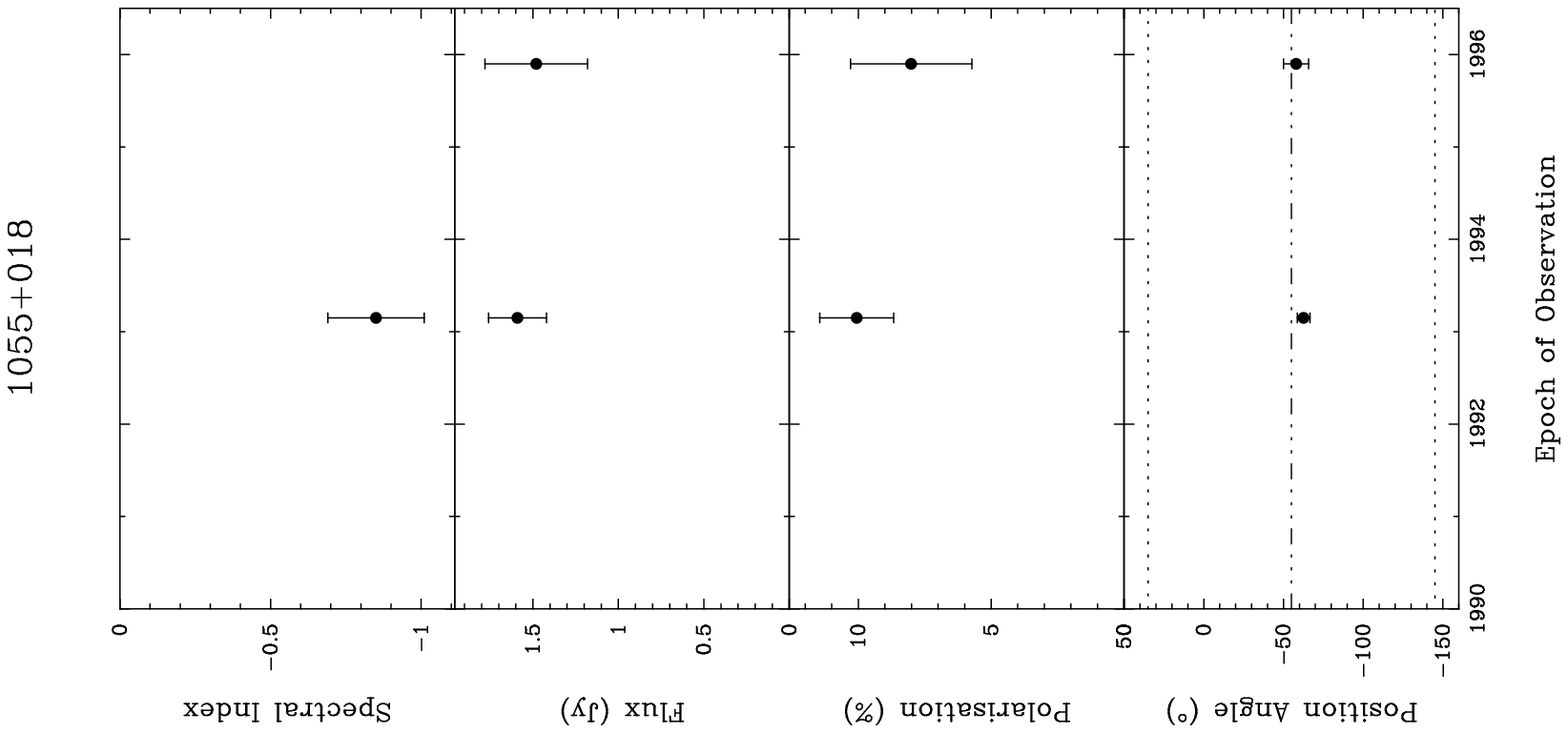} 
 \includegraphics{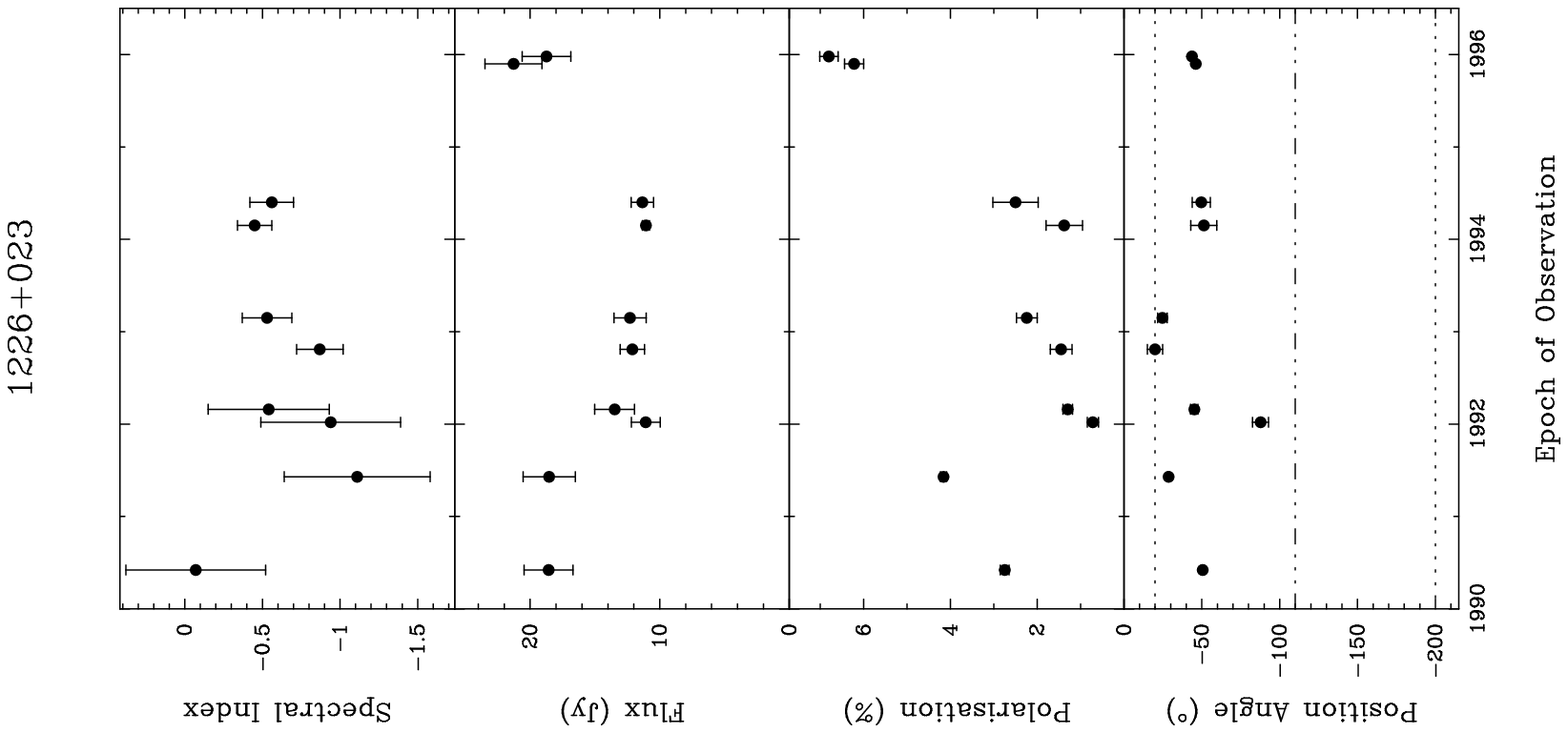} 
 \vspace{1.5cm}
 \contcaption{}
\end{figure*}

\begin{figure*}
 \vspace{10.15cm}
 \includegraphics{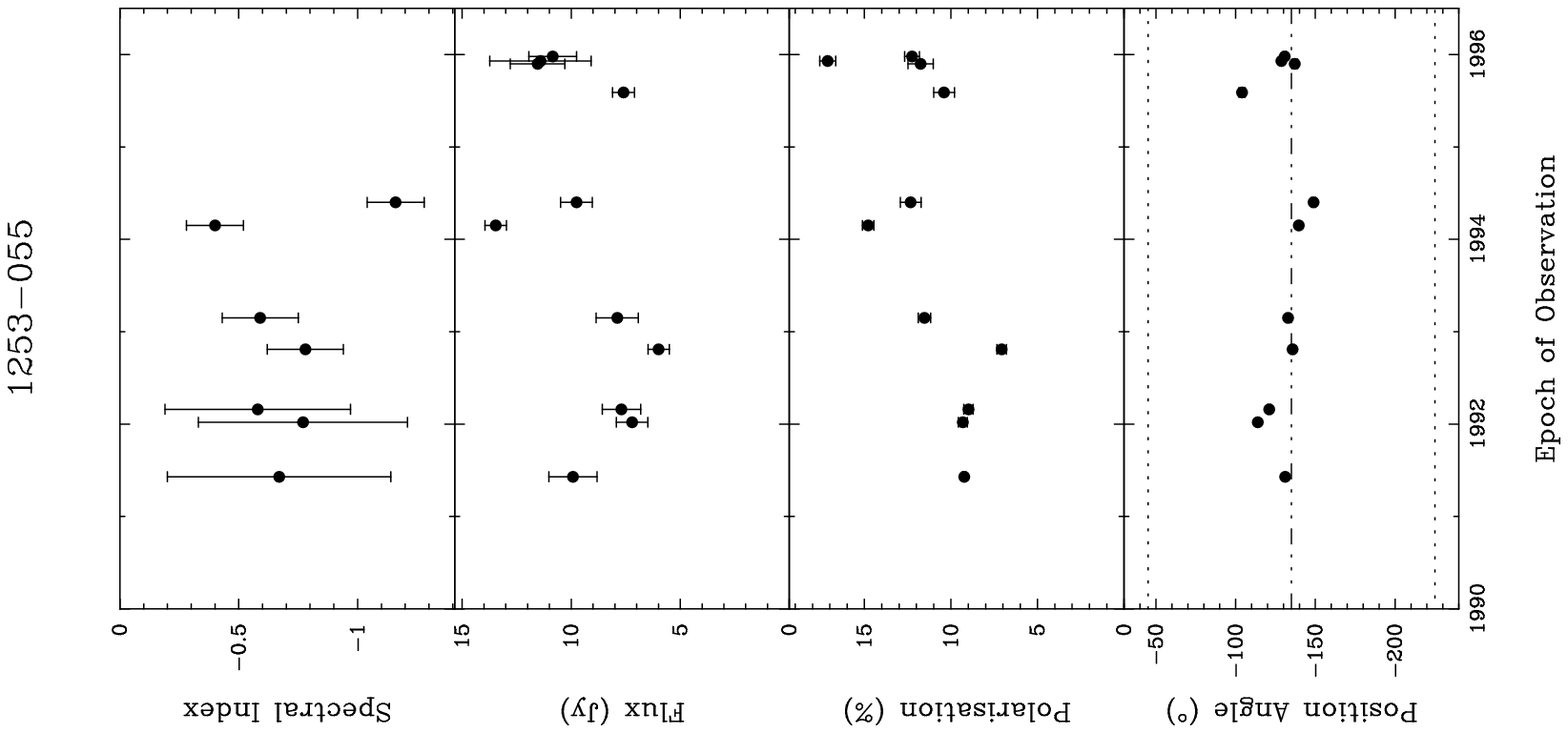} 
 \includegraphics{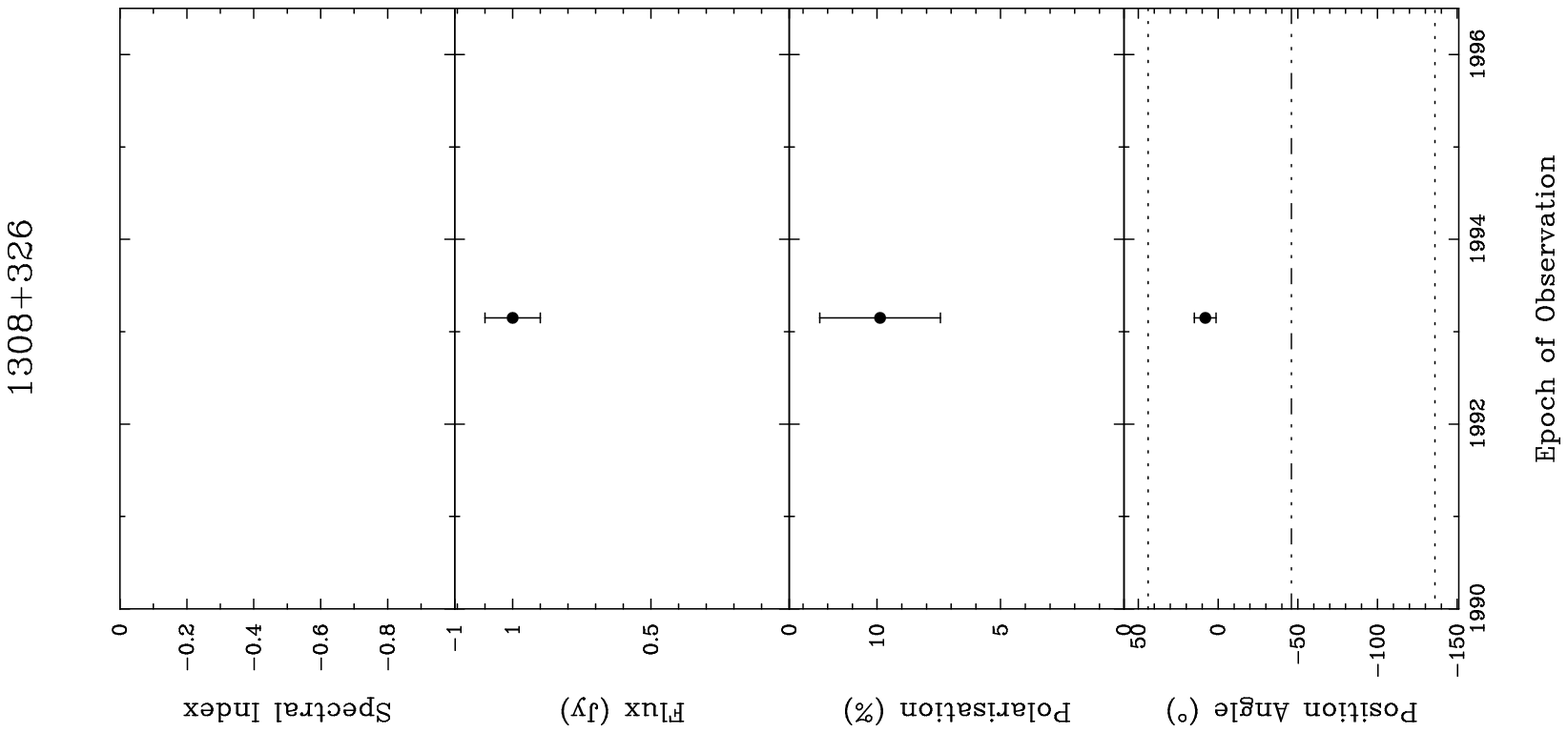} 
 \includegraphics{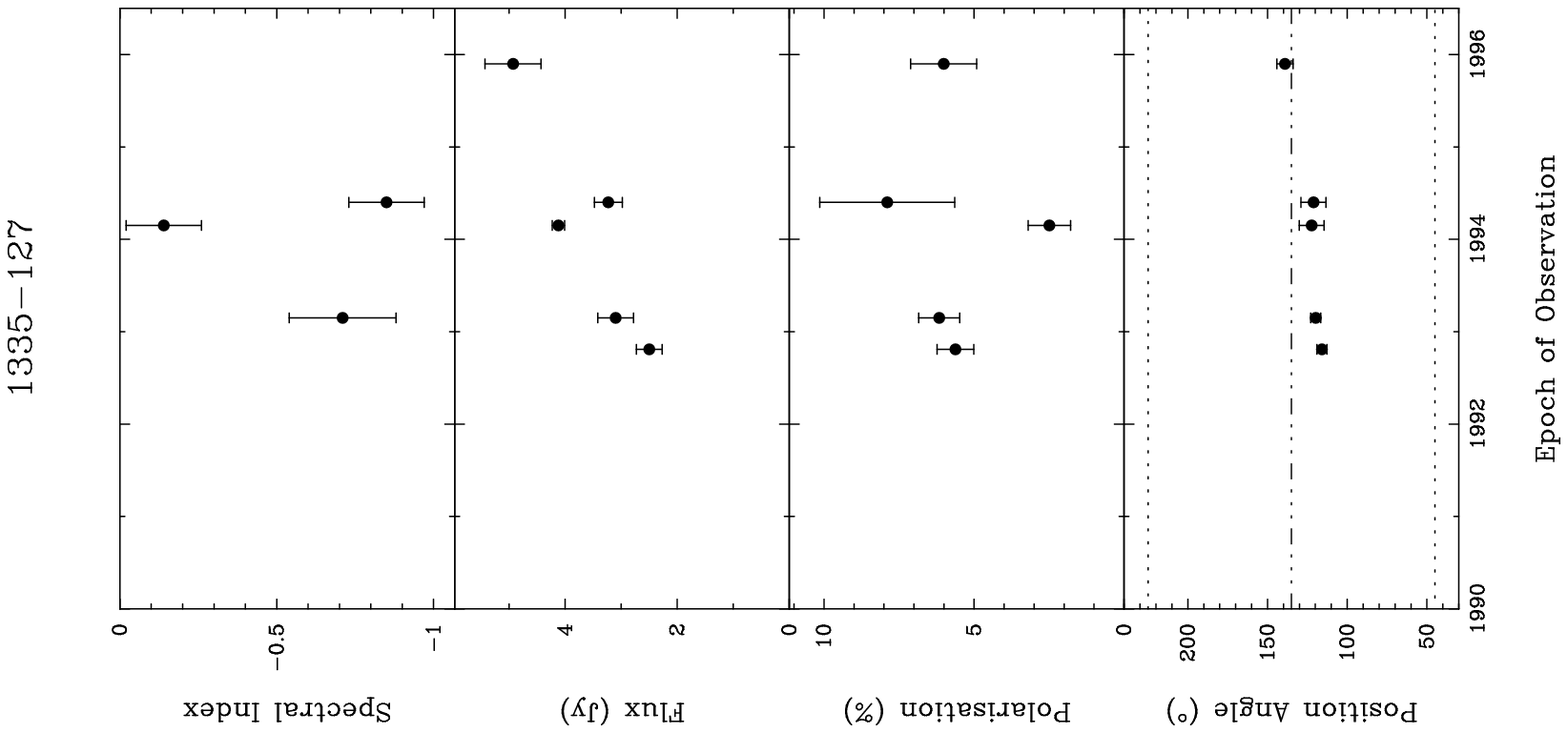} 
 \vspace{11.75cm}
 \includegraphics{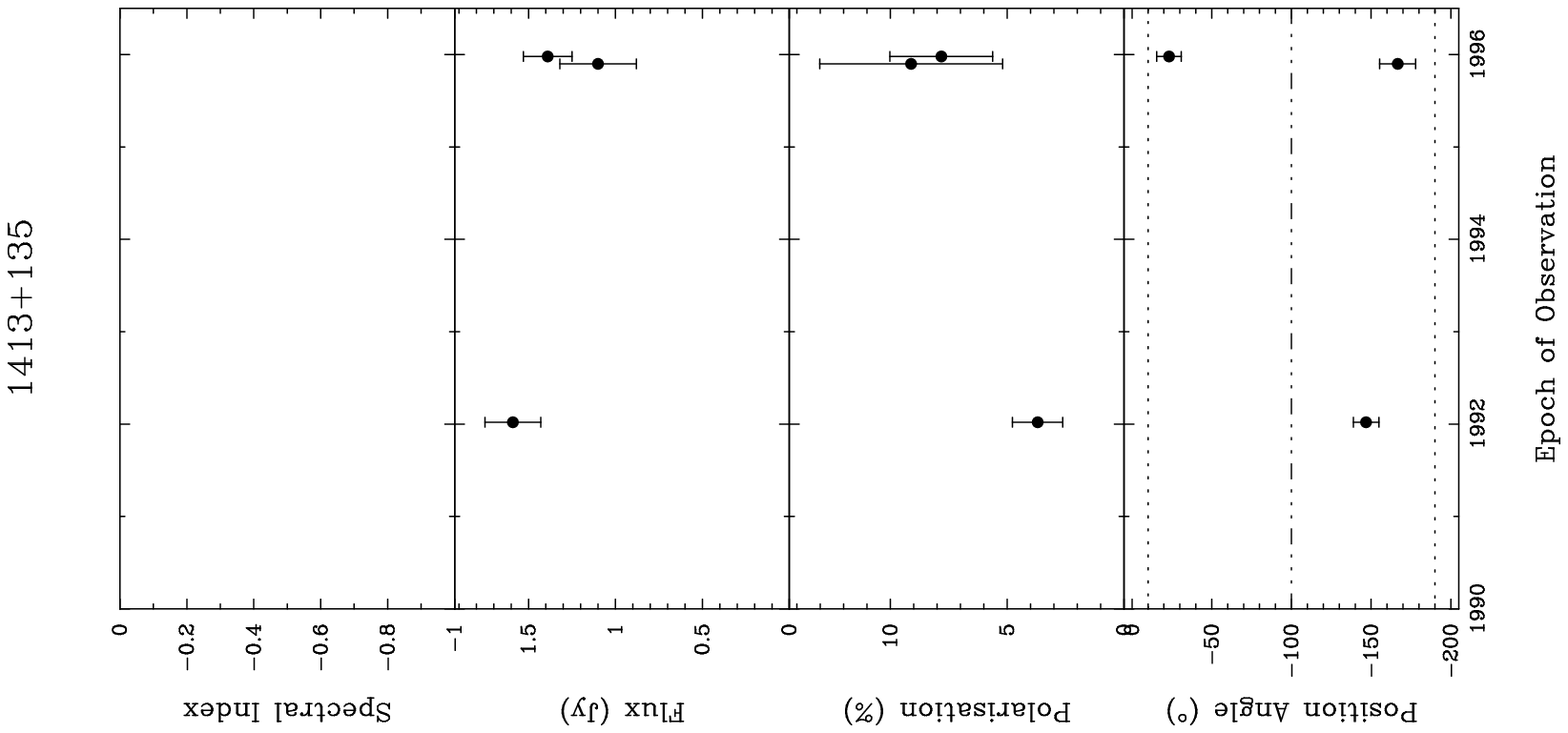} 
 \includegraphics{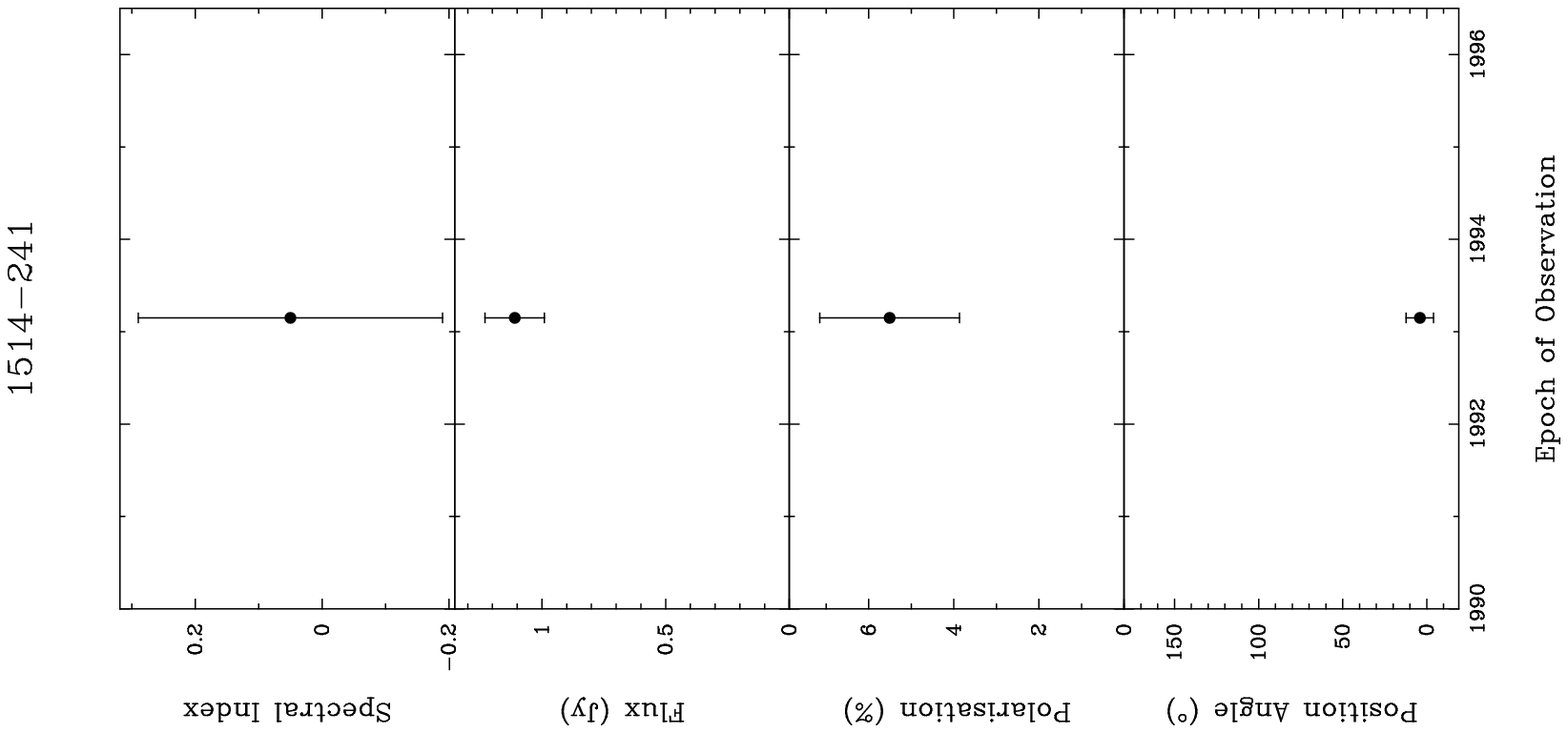} 
 \includegraphics{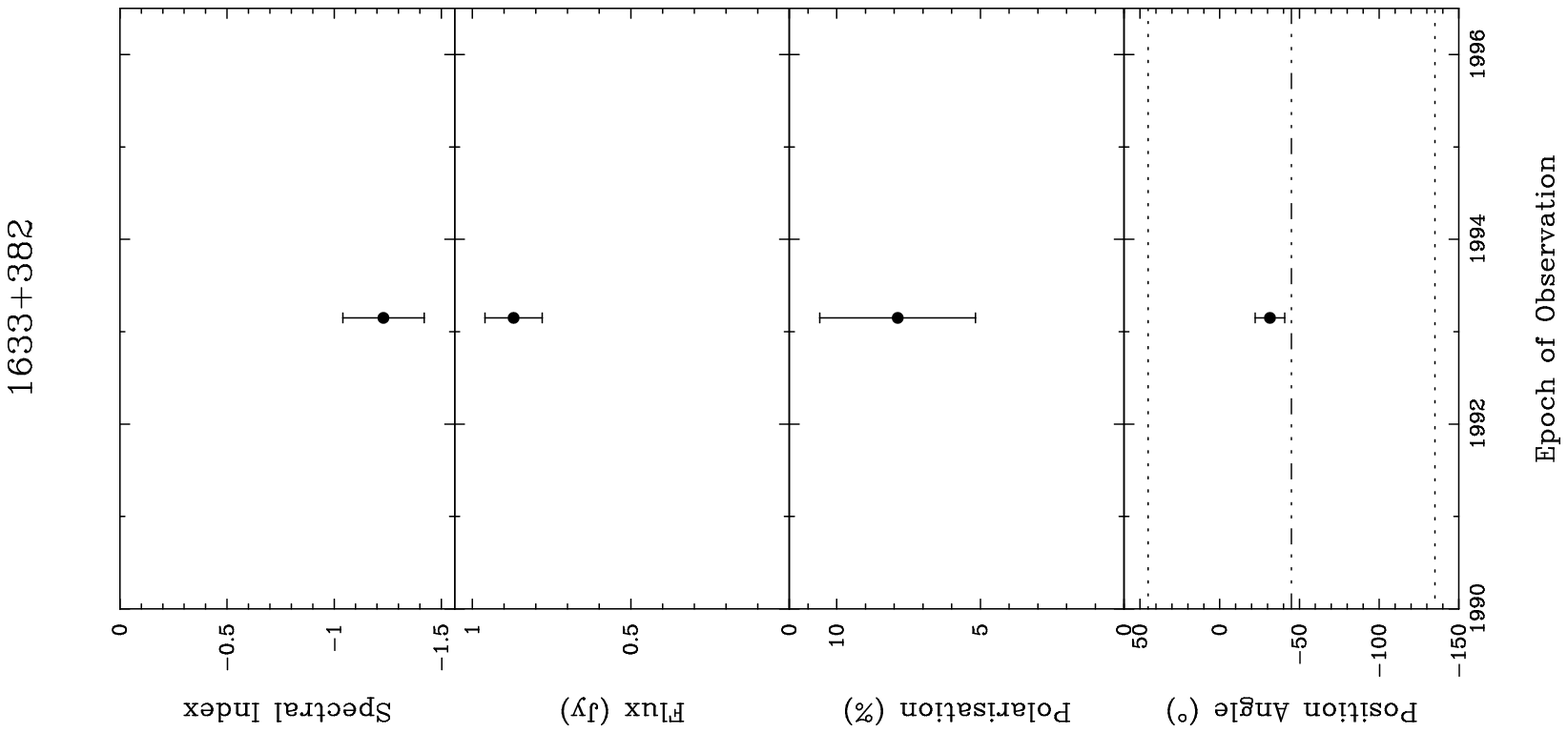} 
 \vspace{1.5cm}
 \contcaption{}
\end{figure*}

\begin{figure*}
 \vspace{10.15cm}
 \includegraphics{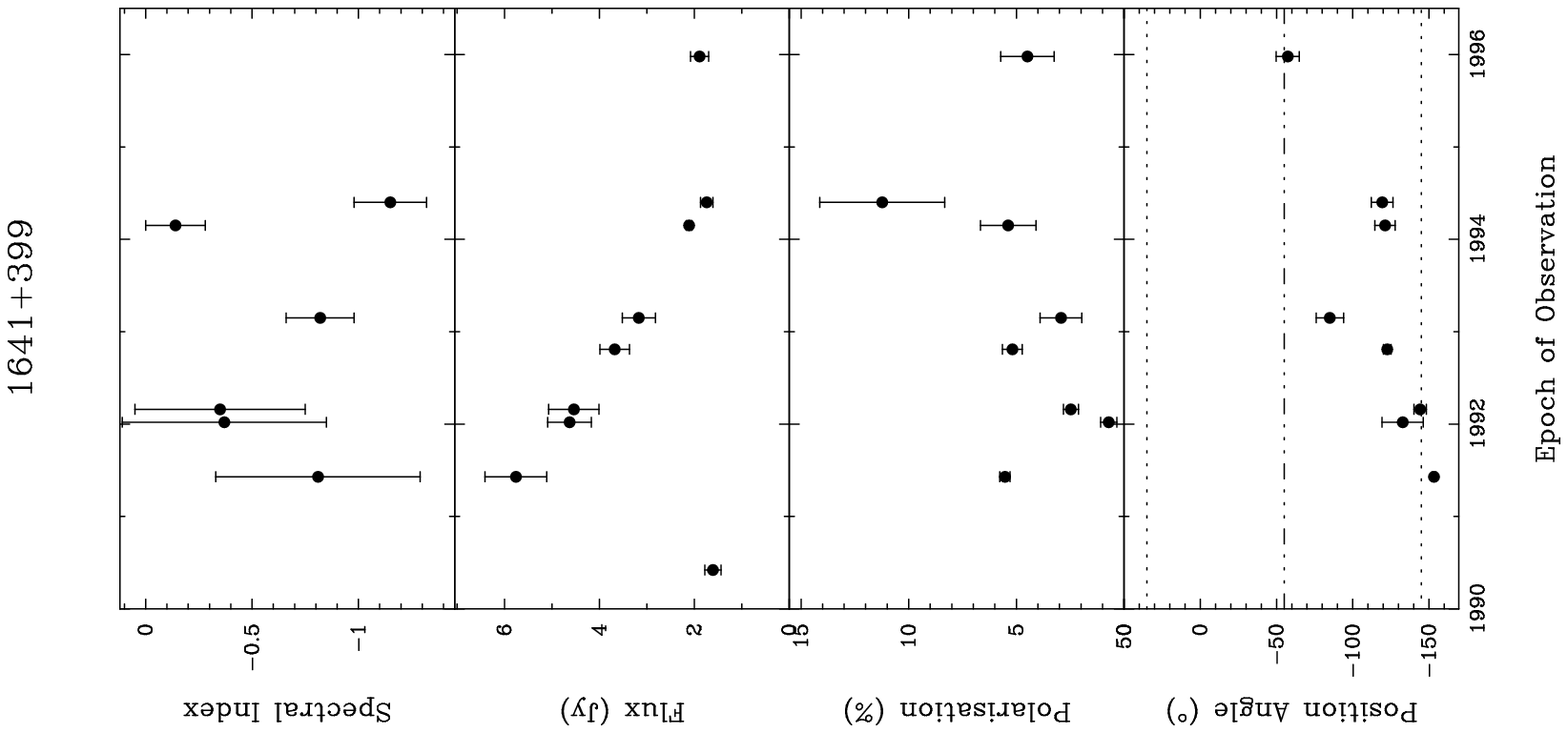} 
 \includegraphics{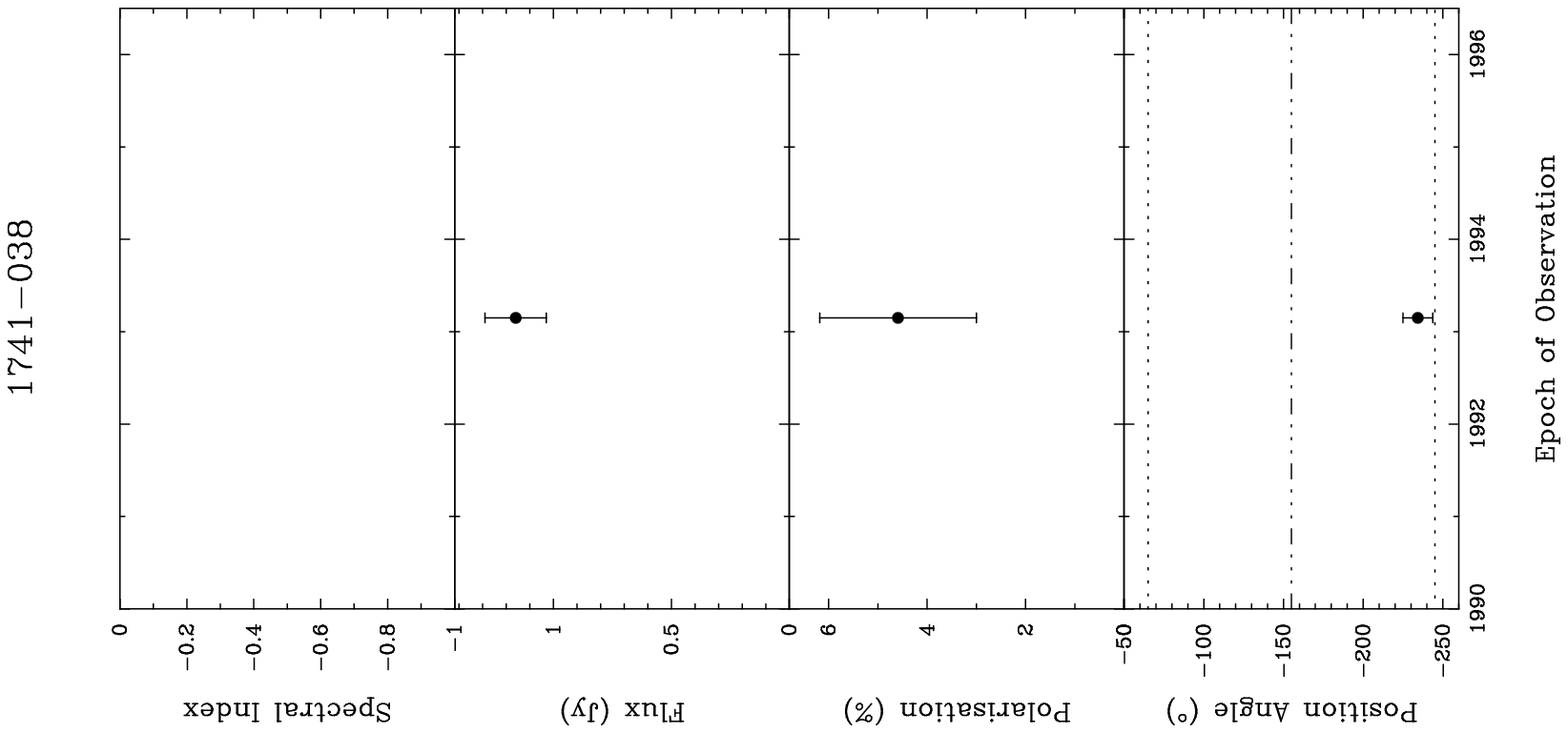} 
 \includegraphics{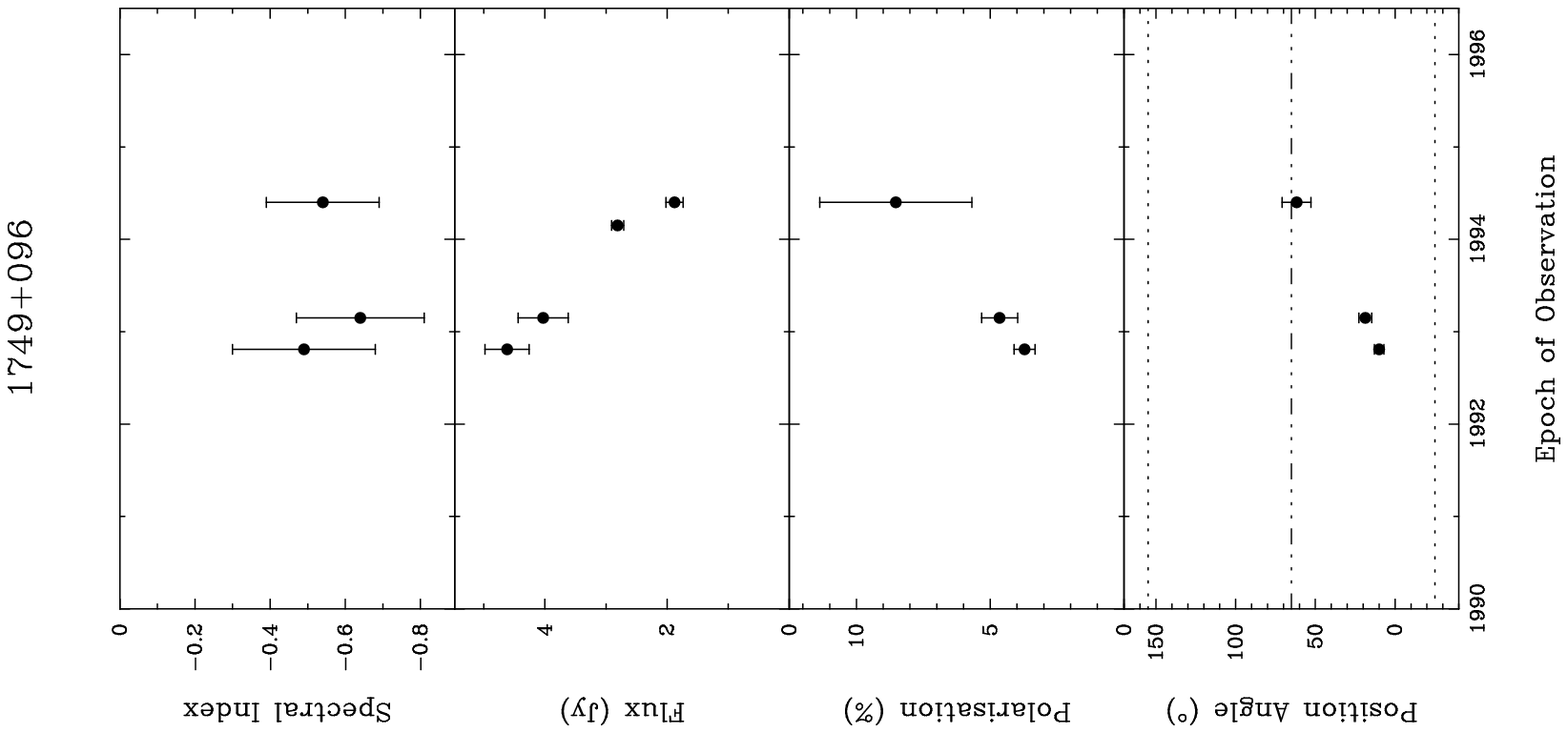} 
 \vspace{11.75cm}
 \includegraphics{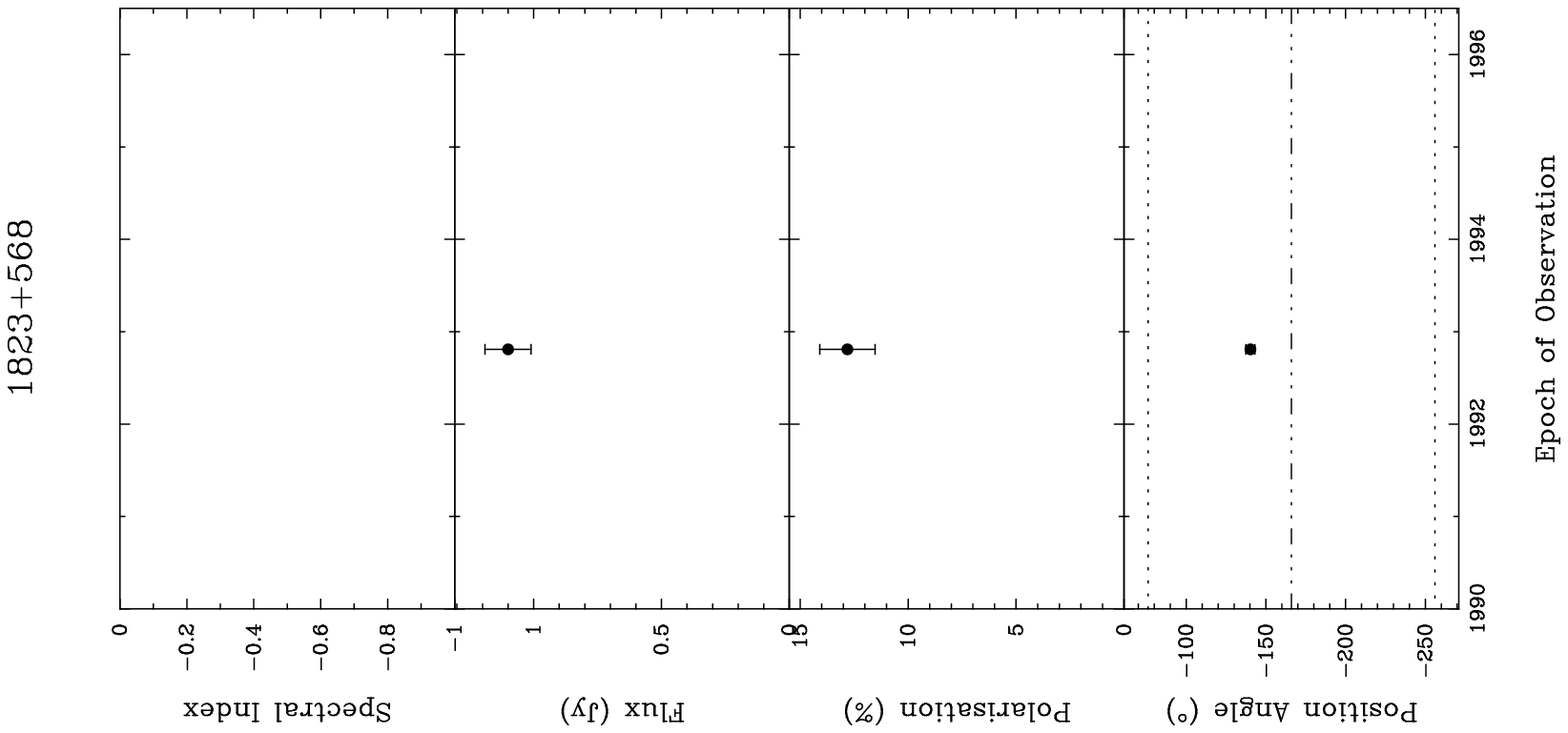} 
 \includegraphics{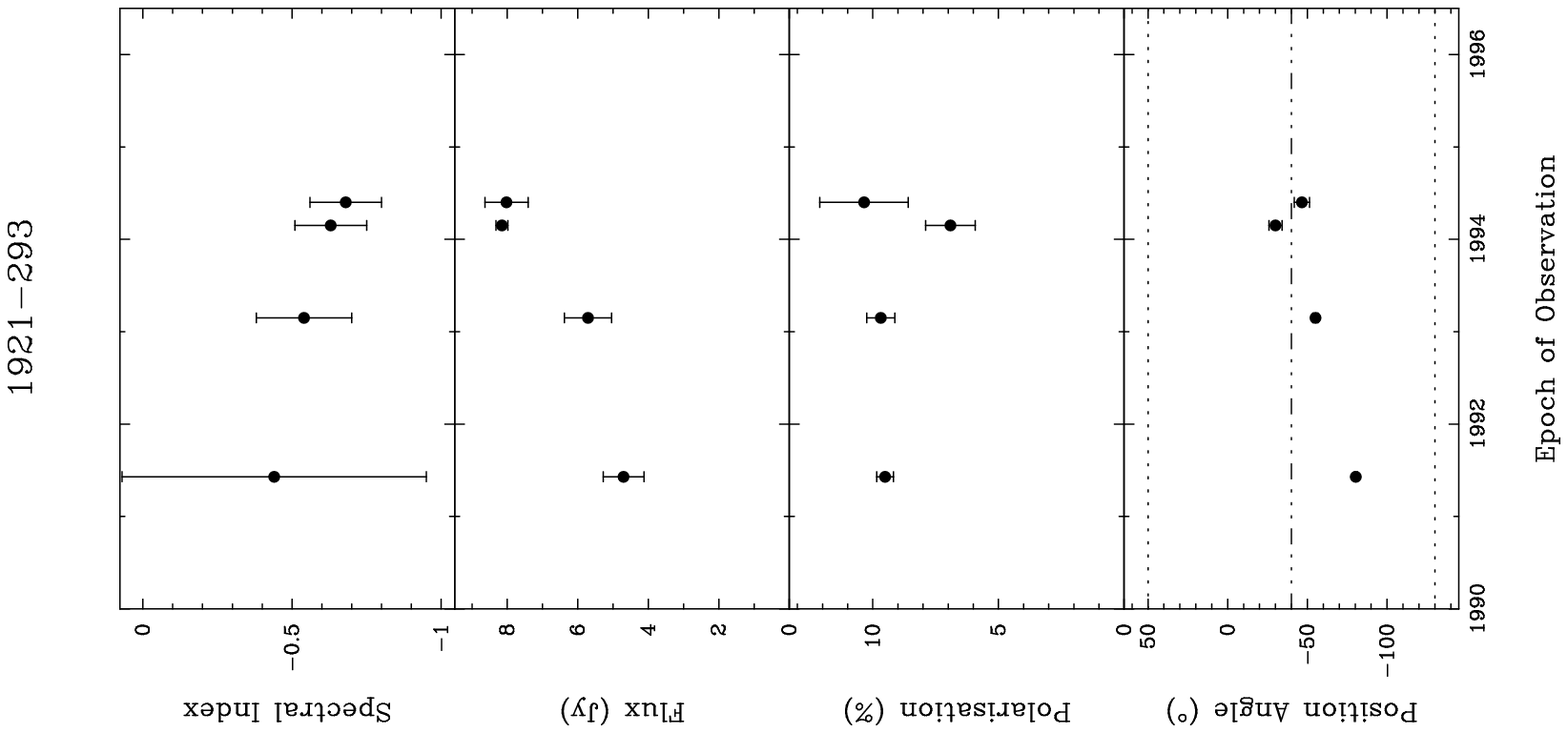} 
 \includegraphics{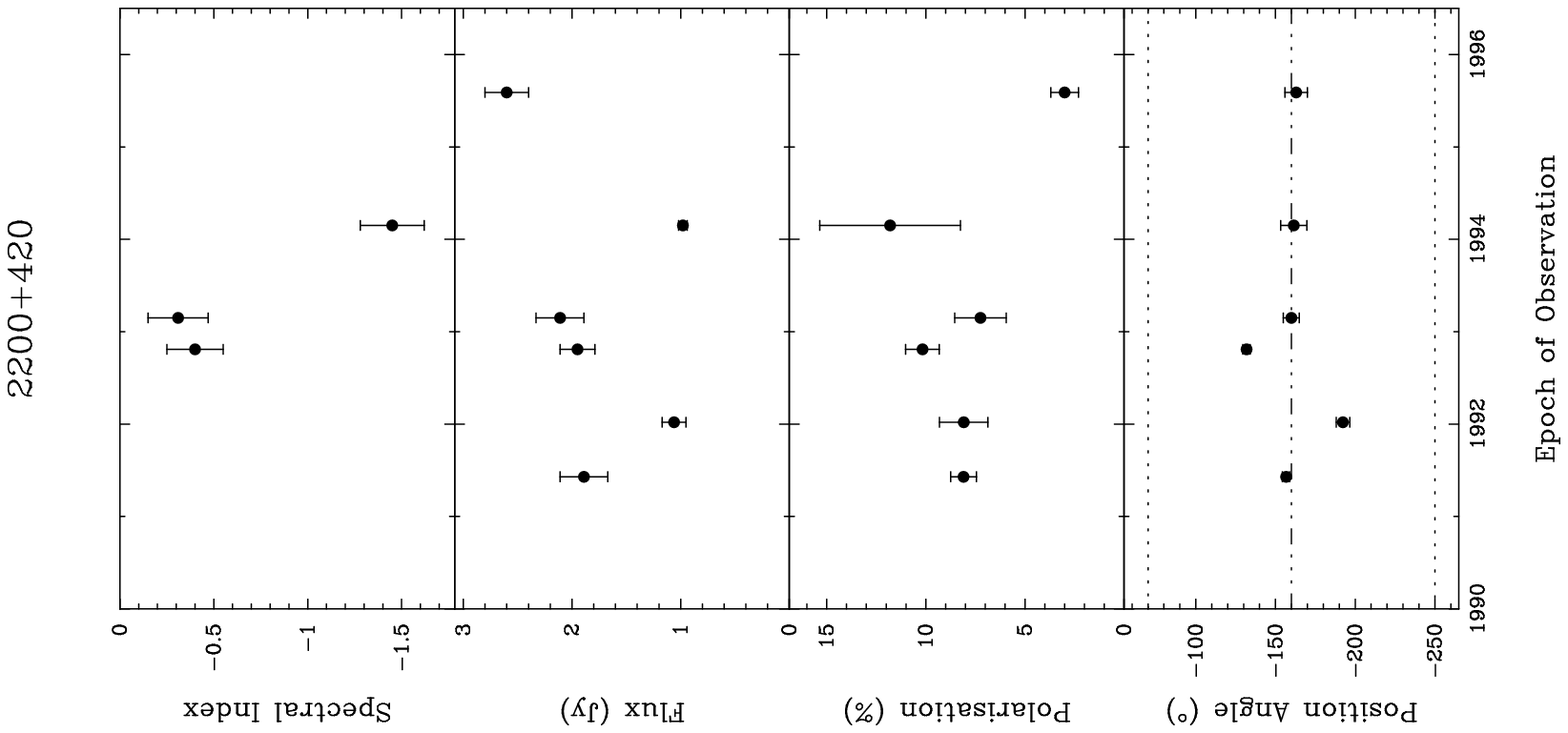} 
 \vspace{1.5cm}
 \contcaption{}
\end{figure*}

\begin{figure*}
 \vspace{10.15cm}
 \includegraphics{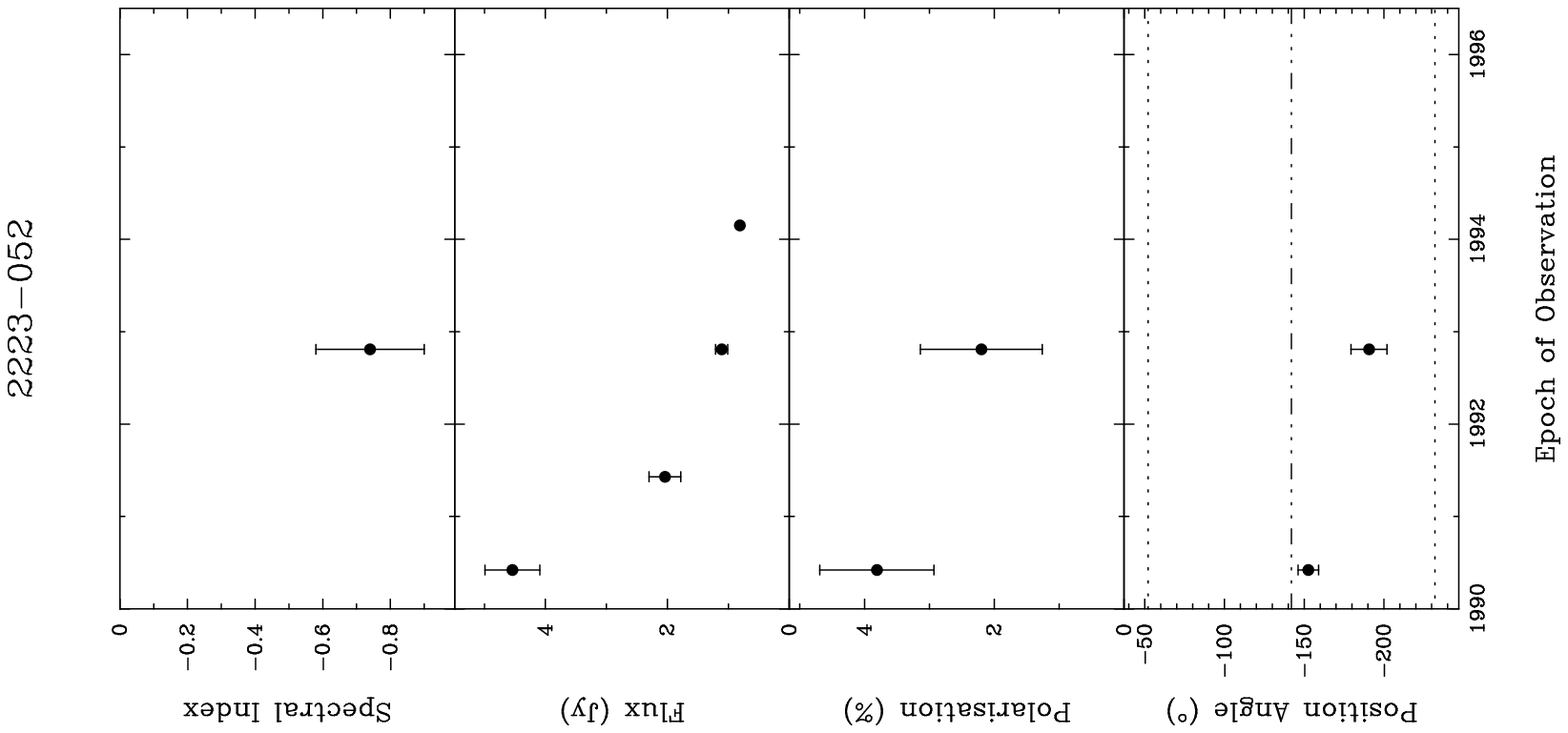} 
 \includegraphics{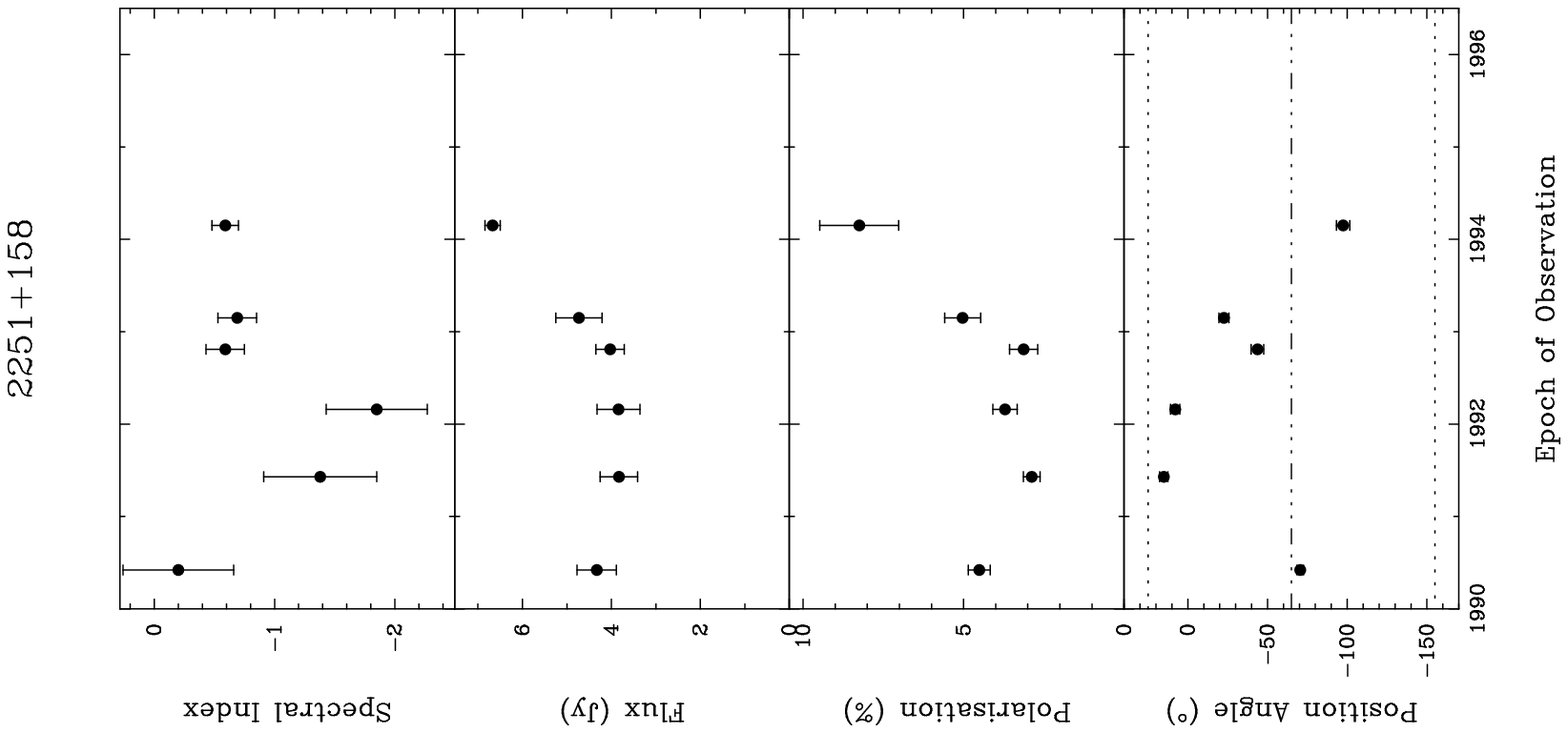} 
 \vspace{1.6cm}
 \contcaption{}
 \vspace{0.9cm}
\end{figure*}

\subsection{Measurements of Spectral Index as a Guide to Opacity}
Photometric data acquired at three millimetre wavelengths have been used to estimate the spectral index.  A flat spectral index suggests that the emission is optically thick, especially if it coincides with a significant increase in the total flux.  If there is significant opacity, synchrotron self-absorption can swing the polarisation position angle by $90\degr$ (Pacholczyk 1970).  There are 12 sources with at least three epochs of simultaneous measurements of $\alpha$, $F$, $P$ and $\chi$ (see data in Table~1 and Figure~3) which allow us to test for opacity.  

The variability in the spectral index measurements is different between sources.  In objects 0235+164, OJ287, 1749+096 and 1921$-$293, the spectral index varies within a small range of values $-0.7 < \alpha < -0.3$, despite significant changes in $F$, $P$ and $\chi$.  In 3C454.3, however, the spectral index takes values $-1.8 < \alpha < -0.2$, and although position angle swings by as much as $90\degr$ have also been observed, neither of these changes can be associated with `flares' in the source emission, as the flux remained constant over the first five epochs and the polarisation level varied only slightly.  We do find a tendency for a flatter spectral index as the flux increases in 3C279, 1335$-$127 and BL Lacertae, but while the polarisation also increases in 3C279, it decreases in the other two sources.  In all three objects, however, $\chi$ remains largely unchanged, which is inconsistent with the emission becoming optically thick during a flare.  In 0420$-$014 the epoch with flatest spectral index coincides with the highest flux and polarisation measured on this source.  In 0735+178, on the other hand, the lowest $\alpha$ is measured when $F$ and $P$ are both low.  In other sources the behaviour is even more complex.  Take for example the last three epochs of 3C345 with spectral index information, obtained in just over a year (between February 1993 and May 1994): the flux, that had been dropping steadily since 1991, falls from 3.2 to 1.7 Jy, yet $\alpha$ changes from $-0.82$ to $-0.14$ to $-1.15$ and $P$ increases from 2.9 to 5.4 to 11.2\%.  These drastic changes in spectral index do not seem to be directly related to changes in flux.  3C273 also exhibits complex behaviour and, although half of the spectral index values available were obtained from two wavelength measurements and therefore have large uncertainties, we have good signal to noise in all $F$ and $P$ measurements.  We have data covering two flaring events during which the polarisation was high.  The epoch of lowest polarisation (0.7\%) and largest position angle swing ($60\degr$), coincides with the lowest flux measured (11 Jy) and a steep spectral index of $-0.94 \pm 0.45$.  The measurements of the less frequently observed sources in the sample all give equally steep spectral indices, with the possible exception of 1514$-$241 ($\alpha = 0.05$).

None of this evidence is suggestive of optically thick emission and we therefore make the asumption that all sources have optically thin synchrotron spectra.

\subsection{Degree of Polarisation}
Examination of the database shows that the degree of polarisation {\em P} can vary considerably in most sources, even in timescales of a few weeks (see e.g. the change in the 1.1 mm percentage polarisation in sources 0420$-$014, 0735+178, OJ287, 3C273 \& 3C345 between epochs 1992.02 and 1992.16 given in Table~1).  However, there seems to be a characteristic range of polarisation values for each source: for example in 3C273 $0.7\% < P < 6.8\%$, in 3C279 $7\% < P < 17\%$ and in 3C454.3 $3\% < P < 8.5$.  We have estimated the {\em mean} polarisation $\bar{P}$ for 20 different sources by averaging the measurements over all epochs available for each object.  There are another 6 sources measured only once, and 2 further objects on which we have $1\sigma$ upper limits.  The polarisation values obtained are listed in Table~2.  

\begin{figure}
 \vspace{6.5cm}
 \includegraphics{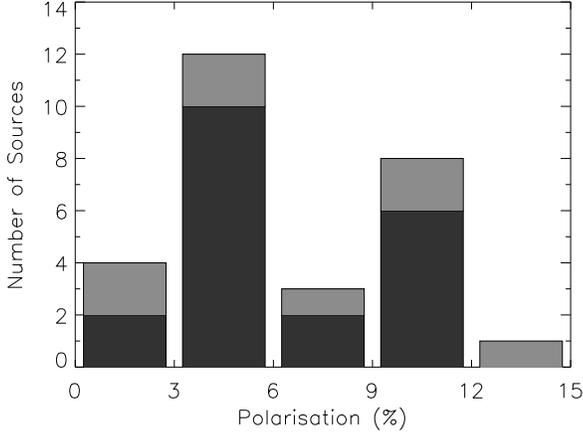}
 \caption{Distribution of percentage polarisation plotted in 3\% bins.  The dark bars represent the {\em mean} polarisation values
          obtained on sources with multiple observations and the light bars show the data on single-epoch sources.}
\end{figure}

In particular, we find that the millimetre emission from OJ287, 3C279, 1921-293 and BL Lacertae is {\em always} highly polarised, with $\bar{P}$ in the range $9 - 11\%$ in all cases, while that from 3C273 and 3C84 typically remains at much lower levels ($2 - 3\%$ or less).  For a majority of sources, though, the mean polarisation level is typically $4 - 7\%$.  Figure~4 shows the distribution of $\bar{P}$ for multi-epoch sources (dark bars), to which we have added the measurements from single-epoch sources (shaded bars).  The main peak of this distribution is $\sim 4.5\%$ and there is a strong secondary peak at $\sim 10\%$.  

We show below that these two broad groups of sources with `high' and `intermediate' polarisation levels behave somewhat differently in terms of the changes in the polarisation position angles measured.  These data also show that although the magnetic fields are certainly ordered on sub-parsec scales, where the millimetre emission is coming from, the degree of ordering is much less than further out in the jets, where polarisation-sensitive VLBI has revealed regions with polarisation levels close to the theoretical maximum for synchrotron radiation ($\sim 70\%$).  It is also possible, however, that the fields are ordered on even smaller scales and that the polarisation detected is the average of several different regions.  Only polarisation-sensitive VLBI at 1mm would unambiguously settle this question.

\begin{table*}
 \caption{Source name, optical type, mean percentage polarisation and VLBI information is given for all sources detected.  Column 1 lists the sources observed, some with alternative names commonly used.  Columns 2 and 3 give the optical identification and references in terms of the following classes: BL Lac, HPQ (High Polarisation Quasar), LPQ (Low Polarisation Quasar) and RG (Radio Galaxy).  Columns 4 and 5 are the mean percentage polarisation at 1.1 mm and standard deviation for sources with multi-epoch observations.  There are eight single-epoch sources (two of which are $1\sigma$ upper limits) for which the polarisation measured and its error are quoted.  The number of polarimetric epochs available on each source is given in column 6.  Columns 7 $-$ 11 give the position angle of the jet as determined from VLBI maps, the distance from the core, frequency of the VLBI observation, references and comments.}

 \begin{center}
  \begin{tabular}{lccrrrrrccl} 
   \cline{1-11}
   \\ \multicolumn{1}{c}{\bf Source Name} & \multicolumn{2}{c}{\bf Optical ID} & \multicolumn{3}{c}{\bf Polarisation} 
      & {\bf $\theta_{\scriptscriptstyle VLBI}^{\circ}$} & $r_{\scriptscriptstyle mas}$ & {\bf GHz} & {\bf Refs.} & {\bf Comments} \\
      & {\em Class} & Ref. & \% & $\sigma$ & n &&&&&  
   \\
   0133+476            &  HPQ  & [c]     &  10.5 &  1.8 &  1 &   165  &    1.3  &   5 & [19] & Low resolution map \\
   0235+164            & BLLac & [b],[g] &   4.9 &  1.0 &  4 & $-150$ &    0.5  &  22 &  [6] & Weak component \\
   0316+413     (3C84) &   RG  & [b],[g] &   1.4 &  0.5 &  2 & $-150$ & $< 1.0$ &  43, 100 & [2],[9] & High resolution maps \\
   0415+379    (3C111) &   RG  &         &  ${\scriptstyle \leq} 0.7$ &&  1 &&  &     &      & No VLBI maps available \\
   0420$-$014          &  HPQ  & [e],[g] &   5.0 &  0.6 &  4 & $-150$ & $< 2.0$ &  43 & [7],[12] & Two independent maps \\
   0430+052    (3C120) &   RG  & [a],[g] &  ${\scriptstyle \leq} 2.2$ &&  1 & $-100$ & $< 20.0$&   5 & [15] & Low resolution, compact core \\
   0528+134            &  LPQ  & [c]     &   3.9 &  1.0 &  2 &    70  &    0.3  &  22 & [21] & PA nearest to core \\
   0735+178            & BLLac & [b],[g] &   6.3 &  1.3 &  4 &    73  &    0.4  &  22 & [20] & Same PA at lower frequencies  \\
   0736+017            &  HPQ  & [c],[e] &   9.8 &  4.6 &  2 &  $-80$ &         &  22 &  [3] & No other maps available \\
   0829+046            & BLLac & [a],[b] &   3.9 &  1.0 &  2 &        &         &     &      & No VLBI maps available \\
   0851+202    (OJ287) & BLLac & [b],[g] &  10.5 &  1.4 &  8 & $-120$ & $< 0.1$ & 43, 100 &  [2],[9] & PA similar to that at 5 GHz \\
   0923+392  (4C39.25) &  LPQ  & [c]     &   4.7 &  1.4 &  4 &  $-88$ &    2.2  &  43 &  [1] & Unresolved core  \\
   1055+018            &  HPQ  & [b],[c] &   9.1 &  1.0 &  2 &  $-55$ &    8.0  & 1.7 & [13] & Low frequency maps \\
   1226+023    (3C273) &  LPQ  & [b],[c] &   2.1 &  0.4 & 10 & $-110$ & $< 0.3$ & 43, 100 &  [2],[8] & Several maps available\\
   1253$-$055  (3C279) &  HPQ  & [e],[c] &  10.5 &  1.0 & 11 & $-135$ & $< 0.1$ & 100 &  [2] & Well determined PA \\
   1308+326            & BLLac & [b],[d] &   9.9 &  2.4 &  1 &  $-46$ & $< 0.7$ &   5 &  [5] & Single VLBI observation \\
   1335$-$127          &  HPQ  & [b],[c] &   5.5 &  1.3 &  5 &   135  & $< 0.2$ &   5 & [18] & Core unresolved \\
   1413+134            & BLLac & [a]     &   6.9 &  2.3 &  3 & $-100$ & $< 8.0$ & 8.4 & [14] & Low resolution \\
   1514$-$241          & BLLac & [a]     &   5.5 &  1.6 &  1 &        &         &     &      & No VLBI maps available \\
   1633+382            &  LPQ  & [c],[g] &   7.9 &  2.7 &  1 &  $-45$ &    1.3  &   5 &  [4] & PA of inner jet \\
   1641+399    (3C345) &  HPQ  & [c],[e] &   4.8 &  1.4 &  8 &  $-55$ & $< 0.2$ & 43, 100 & [2][10][16] & Large range of PAs \\
   1741$-$038          &  HPQ  & [c]     &   4.6 &  1.6 &  1 & $-155$ &         &  22 & [12] & Short, weak jet \\
   1749+096            & BLLac & [a],[b] &   5.6 &  1.8 &  3 &    65  & $\sim 1.0$ &   5 &  [18] & Unresolved, East elongation \\
   1823+568            & BLLac & [d]     &  12.8 &  1.3 &  1 & $-166$ &    0.8  &   5 &  [4] & Agrees with earlier maps \\
   1921$-$293          &  HPQ  & [b],[g] &   9.1 &  0.9 &  4 &  $-40$ &         &  43 & [17] & No other maps available \\
   2200+420   (BL Lac) & BLLac & [b],[g] &   9.1 &  1.2 &  6 & $-160$ & $< 1.5$ & 100 &  [2] & Same PA at lower frequencies \\
   2223$-$052  (3C446) & BLLac & [f],[g] &   3.0 &  1.0 &  2 & $-142$ & $\sim 0.1$ & 100 & [11] & PA of innermost jet \\
   2251+158  (3C454.3) &  HPQ  & [c],[e] &   4.6 &  0.8 &  6 &  $-65$ &   0.6   &  43 &  [7] & Same PA at lower frequencies \\
   \\ \cline{1-11}
  \end{tabular}
 \end{center}

\vspace{1mm}

\begin{flushleft}
{\scriptsize OPTICAL ID REFERENCES $-$ [a] Burbidge \& Hewitt (1992); [b] Ghisellini {\em et al.} (1993); [c] Impey \& Tapia (1990); [d] K\"{u}hr \& Schmidt (1990); [e] Moore \& Stockman (1984); [f] Rusk (1990); [g] Wehrle {\em et al.} (1992).}

\vspace{1.5mm}

{\scriptsize VLBI REFERENCES $-$ [1] Alberdi {\em et al.} (1993); [2] B\"{a}\"{a}th {\em et al.} (1992); [3] Bloom {\em et al.} (1996); [4] Cawthorne {\em et al.} (1993); [5] Gabuzda {\em et al.} (1992); [6] Jones {\em et al.} (1984); [7] Kemball {\em et al.} (1996); [8] Krichbaum {\em et al.} (1990); [9] Krichbaum {\em et al.} (1992); [10] Krichbaum {\em et al.} (1993); [11] Lerner {\em et al.} (1993); [12] Marscher {\em et al.} (1997); [13] Padrielli {\em et al.} (1986); [14] Perlman {\em et al.} (1994); [15] Pilbratt {\em et al.} (1987); [16] Rantakyr\"{o} {\em et al.} (1992); [17] Shen {\em et al.} (1997); [18] Wehrle {\em et al.} (1992); [19] Zensus {\em et al.} (1984); [20] Zhang \& B\"{a}\"{a}th (1991); [21] Zhang {\em et al.} (1994).}
\end{flushleft}
         
\end{table*}

\subsection{Relationship between Degree of Polarisation and Total Flux}
It is possible to investigate the relationship between the degree of polarisation and total flux of those sources that have been observed at a number of different epochs.  Figure~5 is a plot of these two parameters for all sources with at least three measurements.  For some of the most frequently observed, namely 3C273, 3C279 and 3C454.3, a strong and statistically significant positive linear correlation is found between the polarised and total intensities (with $r \geq 0.8$ and confidence level $\geq 99.4$ per cent).  Very tentative positive correlations between $P$ and $F$ may also occur in 0420$-$014 and $0735+178$.  In other sources, however, there is some evidence that these parameters may be anti-correlated, for example in $0235+164$, 3C345, $1749+096$ and BL Lacertae.  OJ287 is a special case since, although no statistical correlation can be found when all 8 observations available are considered, this is due to a change in the sense of the correlation from positive to negative at an epoch when the flux density is a maximum and the polarisation is at minimum level (see Appendix~A for further details).  In 4C39.25, 1334$-$127 and 1921$-$293, however, there is no clear relationship between these two parameters. 

\begin{figure}
 \vspace{23cm}
 \includegraphics{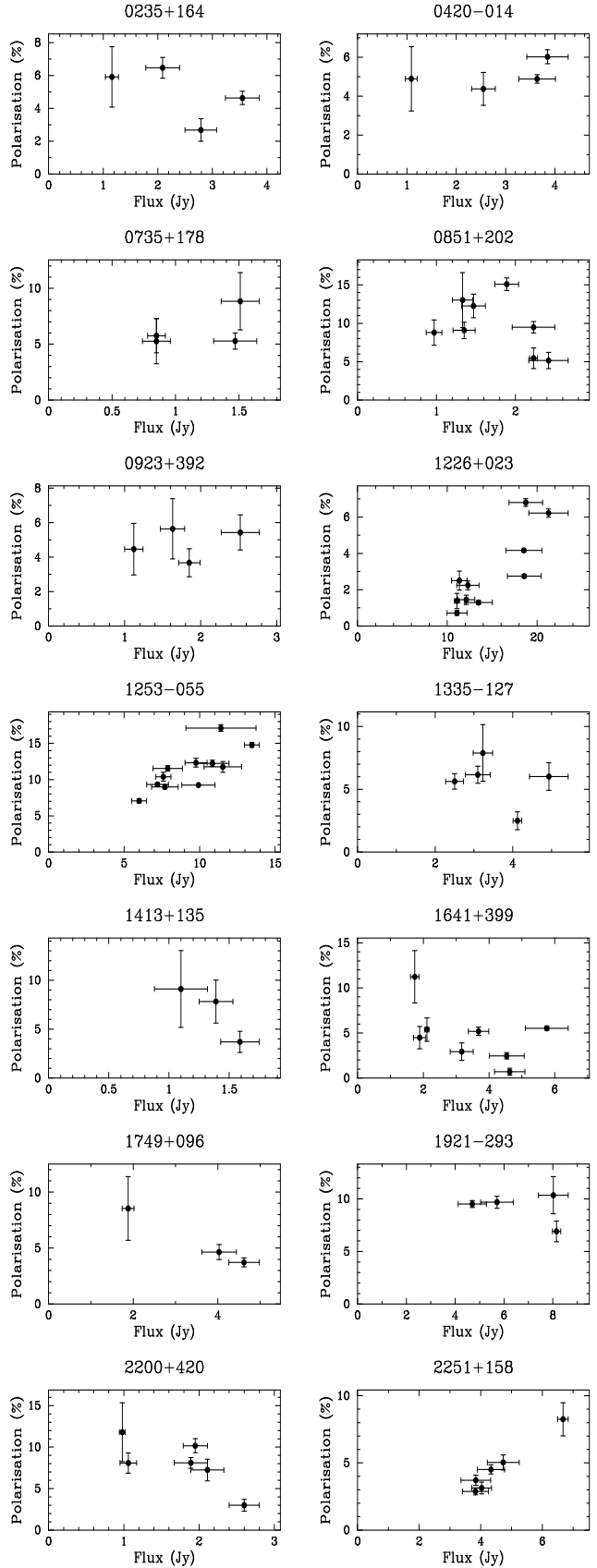}
 \caption{Percentage polarisation is plotted against flux density for all sources on which there are at least three epochs of
          data.  }
\end{figure}

These results can shed light into the physical processes taking place in the most compact regions of the jets and could also provide clues about their morphology, as will be discussed later in $\S 5$.

\subsection{Magnetic Field Orientation}

\begin{figure*}
 \vspace{12.5cm}
 \includegraphics{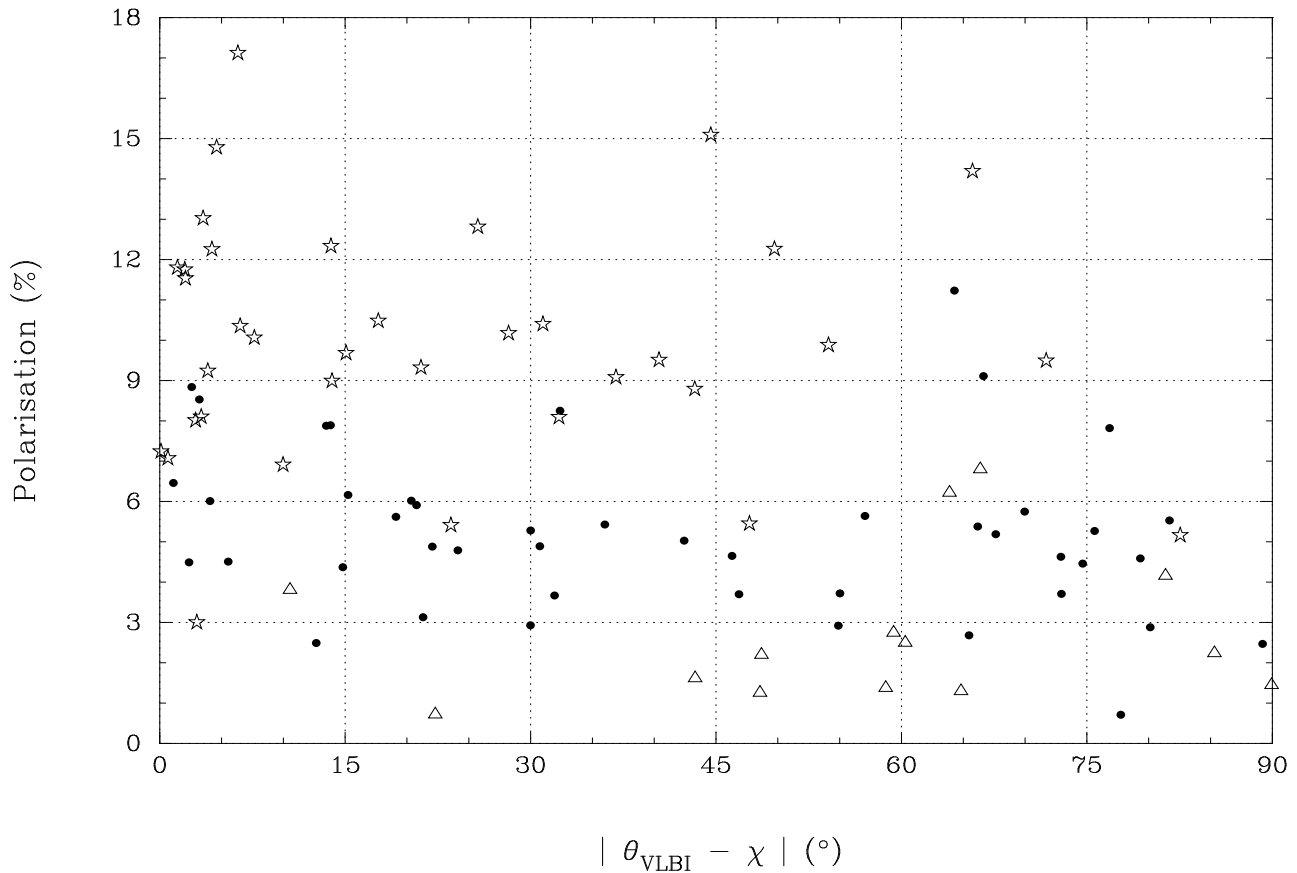}
 \caption{Percentage polarisation as a function of magnetic field orientation with respect to the inner jet direction.  Different 
          symbols have been used to represent measurements of sources whose mean polarisation $\bar{P}$ lie in three different 
          ranges: empty stars for $\bar{P} > 9\%$, solid circles for $9\% > \bar{P} > 3\%$ and empty triangles for 
          $\bar{P} < 3\%$.  Measurements of single-epoch sources have also been included. The error bars have been omitted for
          clarity but all observations are better than $3\sigma$.}
\end{figure*}

In order to relate these polarisation observations to the magnetic field geometry in the source, we make the assumption that the emission is optically thin synchrotron radiation, for which the polarisation position angle is perpendicular to the magnetic field direction in the source frame.  The magnetic field geometry can then be related to the position angle of the jet axis, as determined from VLBI observations.  Many core dominated sources, however, show significant bending from parsec to kiloparsec scales (e.g. see 3C84, 3C273, 3C279, 3C345, 3C454.3 \& 4C39.25 in Readhead {\em et al.} 1983) and we know that if further bending occurs between the scale of the centimetre emission and the millimetre emission, our conclusions could be severely compromised or even completely erroneous.  Even for sources with apparently straight jets from parsec to kiloparsec scales, we cannot rule out the possibility of bending on smaller scales.  Ideally we would like to compare our polarisation data with VLBI maps obtained at 1 mm, but in practice we are foced to estimate the direction of the jet structure from the highest-frequency VLBI map available to us.  These have been listed in Table~2, where we have also included the distance from the core to the nearest component (in {\em mas}), the frequency of the observation and a brief comment on the characteristics of the map.  In the rest of this discussion we refer to the magnetic field orientation thus determined, instead of the measured polarisation position angle $\chi$.  The convention used throughout is that the magnetic field is perpendicular to the VLBI jet when $\mid\theta_{VLBI} - \chi\mid$ is $0\degr$ and parallel at $90\degr$.

In Figure~6 we plot degree of polarisation $P$ against $\mid\theta_{VLBI} - \chi\mid$.  We treat measurements of the same source at different epochs as independent points, eventhough some sources have been observed more frequently than others.  This allows a realistic picture of the overall behaviour to emerge, rather than taking an average of the positon angle measured for a source which might show radically different behaviour at different times.  We have used, however, different symbols to represent sources whose mean polarisation $\bar{P}$ lie in three different ranges as discussed above: empty stars for $\bar{P} > 9\%$, solid circles for $9\% > \bar{P} > 3\%$ and empty triangles for $\bar{P} < 3\%$.  Measurements of 8 sources observed only once have also been included as belonging to one of these three categories.  

This figure clearly shows that the more highly polarised sources are rarely found with polarisation levels lower than 6\% and that there is a strong tendency for them to have magnetic fields largely perpendicular to the jet axis.  Only a sixth of the measurements in this subset of objects have $\mid \theta_{VLBI} - \chi \mid > 45\degr$, and 5 of the `starred' data points close to $45\degr$ belong to observations of OJ287 alone.  

Sources with an `intermediate' level of polarisation do not have a preferred magnetic field orientation.  In fact, many of the most frequently observed sources in this category exhibit magnetic fields either parallel or perperdicular to the jet at different epochs, e.g. 0235+164, 0735+178, and any orientation in between, such as 4C39.25, 3C345 and 3C454.3 where the position angle measured is clearly seen to change gradually (in tens of degrees) rather than in $90\degr$ swings.  However, the magnetic field in 1335$-$127 is clearly perpendicular to the jet and does not change over 5 epochs spanning more than 3 years. 

It is also clear from Figure~6 that the measurements from sources with low mean polarisation are largely associated with magnetic fields closer to the aligned configuration rather than perpendicular to the jet.  However, this must be treated with caution as 10 of these points correspond to observations of 3C273 alone.

All of the examples quoted above can be verified by looking at the raw data plotted in Figure~3.  One and two-dimensional Kolmogorov-Smirnov (K-S) tests of $P$ and $\mid\theta_{VLBI} - \chi\mid$ between any two categories all give high K-S statistics with confidence levels $>99.9$ per cent, confirming that these groups have statistically distinct polarisation properties.

\subsection{Polarisation Properties of BL Lac Objects and Flat-Spectrum Radio Quasars}

\subsubsection{Source Classification}
Although all the sources in the sample have so far been generally treated as `blazars', i.e. compact, flat-spectrum radio sources with highly variable and polarised emission, most objects have traditionally been classified as belonging to one of two optical types, namely BL Lac objects (BL Lacs) and Flat-Spectrum Radio Quasars (FSRQs).  The latter can be divided into High Polarisation Quasars (HPQs) (also known as Optically Violently Variable quasars or OVVs) and Low Polarisation Quasars (LPQs) according to whether their optical polarisation is $>3\%$ (Moore \& Stockman 1981).  The classical distinction between BL Lacs and HPQs is now unclear as emission lines have been observed in many BL Lacs (see e.g. Antonucci 1993 for a discussion).

In Table~2 we have indicated the most common classification for each source, and we shall adopt this classification for the purpose of comparing the properties of these three types.  Of the 28 sources in our sample we have identified 11 BL Lacs, 10 HPQs and 3 LPQs.  However, some ambiguity exists about the correct classification of some objects, especially between the BL Lac and HPQ categories.  In fact, 0133+476 and 1335$-$127 have at times been classified as BL Lacs (Zensus {\em et al.} 1984 and Kuhr \& Schmidt 1990, respectively) and 1308+326, 1823+568 and 3C446 have all been classified as HPQs in a number of studies (e.g. Moore \& Stockman 1981, Impey {\em et al.} 1991,  Burbidge \& Hewitt 1992, Ghisellini {\em et al.} 1993).  Also, the BL Lac object $0235+164$ has been proposed as a candidate micro-lensed quasar (Saust 1992).  Sources 3C84, 3C111 and 3C120 are usually described as radio galaxies.

\subsubsection{Comparison between Different Optical Types}

\begin{figure*}
 \vspace{13.25cm}
 \includegraphics{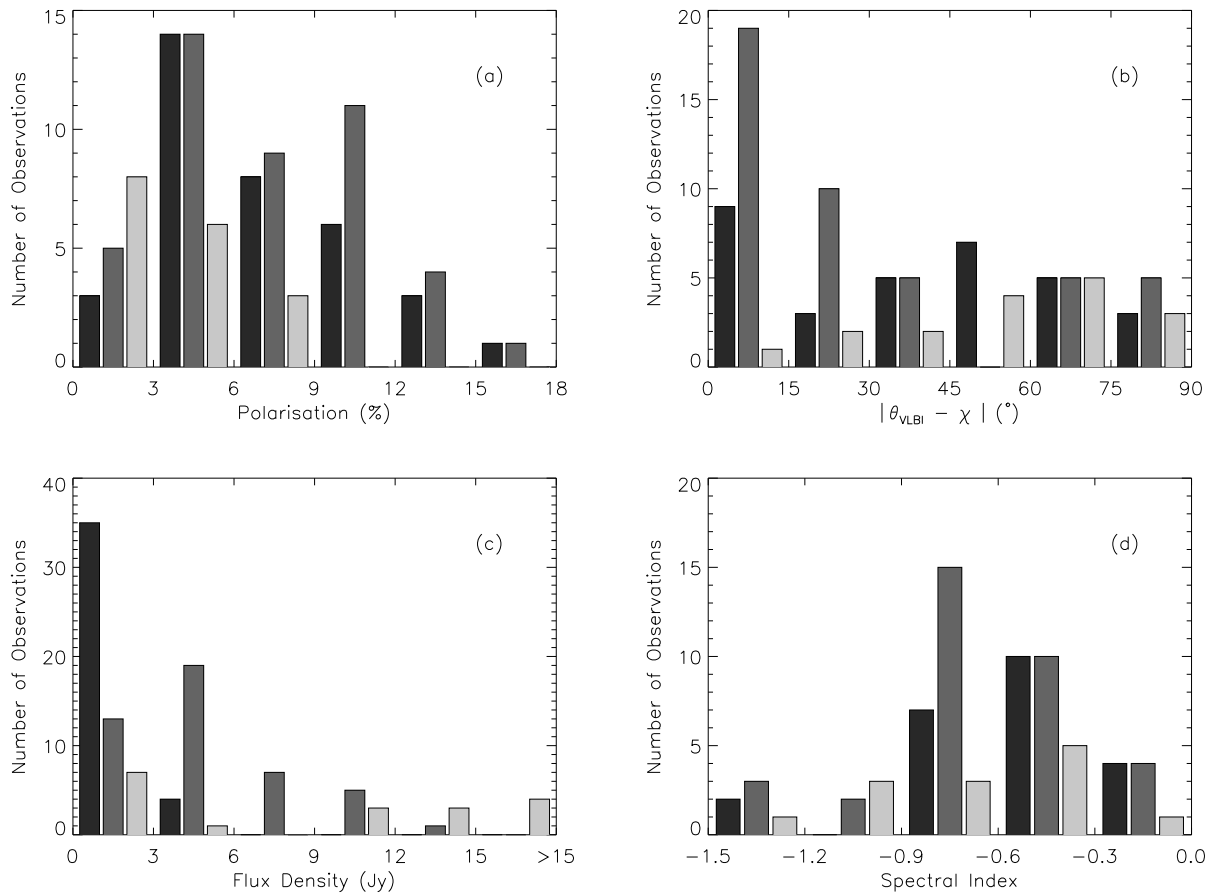}
 \caption{Comparison between the distributions of (a) percentage polarisation, (b) magnetic field orientation, (c) flux density
          and (d) spectral index for the BL Lacs (black bars), the HPQs (dark shaded bars) and the LPQs (light shaded bars).  
          Note that the bin sizes in all four histograms are the same for all three types being compared and are as follows: (a)
          3\%, (b) $15\degr$, (c) 3 Jy and (d) 0.3.  Also note that all the measurements available in the database have been
          included and therefore most sources have multiple entries.}
\vspace{5mm}
\end{figure*}

One of the initial results from the first polarisation-sensitive VLBI surveys at centimetre wavelengths was the claim that BL Lacs and FSRQs differed in the orientation of their magnetic fields with respect to the jet axis, being always perpendicular in the BL Lacs and always parallel in the quasars.  This claim has recently been moderated somewhat, however it is still quite marked on parsec scales (see e.g Cawthorne {\em et al.} 1993 and Gabuzda {\em et al.} 1994).  Our sample contains a similar number of sources of either class (although we shall differentiate between the HPQs and the LPQs amongst the quasars) and there are multiple observations (obtained at different observing epochs) of most objects, making the number of measurements in our database statistically significant.  We are aware, however, that a few sources have been measured more often than others and that this introduces a certain degree of bias; we have tried to point this out wherever it is relevant.

In Figure~7 we compare the distributions of (a) polarisation level, (b) magnetic field orientation, (c) flux density and (d) spectral index for BL Lacs (black bars), HPQs (dark shaded bars) and LPQs (light shaded bars).  The bin sizes of the histograms are indicated by the horizontal axis of the plots: 3\% in $P$, $15\degr$ in $\mid\theta_{VLBI} - \chi\mid$, 3 Jy in $F$ and 0.3 in $\alpha$.

The shape of the distributions of $P$ and $\mid\theta_{VLBI} - \chi\mid$ are `visually' very similar between the BL Lacs and the HPQs.  For both classes, 80 per cent of the polarisation measurements lie in the range $3\% < P < 12\%$ and the magnetic field can take any orientation between the perpendicular and the parallel to the jet axis.  However, while the mesurements of BL Lacs are equally distributed between those with magnetic fields oriented between $0\degr - 45\degr$ and those between $45\degr - 90\degr$, the HPQs favour the perpendicular configuration with a ratio $>$3:1.  In fact, there are no HPQ measurements in the $45\degr - 60\degr$ bin, while 5 different sources contribute to the 7 measurements of BL Lacs in this bin.  We also find contributions from 6 BL Lacs and 7 HPQs in the $0\degr - 15\degr$ bin, and from 6 BL Lacs and 8 HPQs in the $15\degr - 30\degr$ bin.  This shows that the multiple data points from most sources tend to spread throughout the plot.  In contrast, there are 9 measurements of 3C279 (the source most frequently observed) in the first bin alone, 5 measurements of OJ287  between $30\degr - 60\degr$ and 6 measurements of 3C345 between $60\degr - 90\degr$, which influence the concentration of data points in these areas.  The data available on the LPQs (15 measurements) come from 3 different sources, but 3C273 contributes with 10 entries and 4C39.25 with 4.  The distributions of $P$ and $\mid\theta_{VLBI} - \chi\mid$ for this class are therefore strongly biased by the characteristics of a single source and thus any comparison with the other classes must be treated with care.  Although all orientations of the magnetic field have been observed at least once in this group, there is a concentration of measurements close to direction parallel to the jet.  Also, half of the polarisation measurements are in the $< 3\%$ bin and none have been obtained above 8\%.

A K-S test of the $P$ parameter between BL Lacs and HPQs gives a K-S statistic of 0.14 with significance level of 16 per cent, indicating that the two samples being compared could have been obtained from a single population of objects.  A very similar result is obtained on the $\mid\theta_{VLBI} - \chi\mid$ parameter (0.13 with 11 per cent significance) and in a 2-dimensional K-S test combining these two parameters (0.16 with 21 per cent significance).  Similar tests carried out between either of these two classes and the LPQs all give K-S statistics $> 0.5$ with significance levels of $> 99$ per cent, clearly differentiating between them.

We therefore find no evidence of any significant difference in the properties of the magnetic field (degree of order and orientation) between BL Lacs and HPQs on the scales sampled by millimetre/submillimetre measurements.  However, there is some evidence that the magnetic field in the inner jets of LPQs may be generally less-ordered and be preferentially aligned with the jet, although this needs to be confirmed by increasing the number of different LPQs in the sample.

Pannels (c) and (d) of Figure~7 summarise the information obtained from the photometric data.  The distribution of flux density measurements is very different between BL Lacs and HPQs.  Practically all the BL Lac measurements are $<$3 Jy, however the flux distribution for the HPQs peaks between $3 - 6$ Jy and three different sources have been measured with $F > 6$ Jy.  The flux distribution of the LPQs is difficult to asses as it is dominated by the measurements of 3C273.  K-S tests between any two categories all give a statistic of 0.6 with significance level $> 99.9$ per cent, thus differentiating between the three classes.  On the other hand, the distribution of spectral index is similar between BL Lacs and HPQs and the K-S test fails to separate the three classes (with significance level of $<$ 75 per cent in all cases).  However, the BL Lacs measurements average to a somewhat flatter mean spectral index ($-0.56 \pm 0.35$) than the HPQs ($-0.70 \pm 0.36$), which is similar to that of the LPQs ($-0.72 \pm 0.33$), and although the difference in the spectral slope may not be statistically significant, this result is in in very close agreement with that of Gear {\em et al.} (1994) obtained on a larger sample of objects using a greater number of wavelengths.

The relationship between $P$ and $F$ (discussed in $\S 4.3$) could provide very tentative evidence of different behaviour between the BL Lacs and the quasars: in general, $P$ and $F$ seem to have a tendency to be anti-correlated in BL Lacs (e.g. 0235+164, some epochs of OJ287, 0829+046, 1413+135, 1749+096 and especially BL Lacertae), whereas quasars are more likely to show a positive correlation between these two parameters (e.g. 3C273, 3C279, 3C454.4 and perhaps 0420$-$014).  However, there are clear exceptions in either class, namely 0735+178 for the BL Lacs and 3C345 for the quasars.

\section{Discussion}

\subsection{Overview: Shocks-in-Jets}
The most widely accepted model for the energy production and transport mechanisms in compact radio sources considers the injection of material from the nucleus of the AGN into the base of the jet.  As the plasma expands along the jet, flux-freezing implies that the component of the magnetic field perpendicular to the jet axis ($B_{\perp}$) decreases with radial distance as $r^{-1}$, whereas the parallel component ($B_{\parallel}$) should decrease more rapidly as $r^{-2}$ (e.g. Laing 1981).  Therefore, unless the field is being constantly re-isotropised, $B_{\perp}$ should become dominant.  An enhancement of $B_{\perp}$ can also be produced if an injection of relativistic particles into the jet leads to the emergence of a relativistic shock front.  As the shock propagates down the jet, it energises the emitting particles, giving rise to flaring of the emission, and compresses the plasma, thus aligning magnetic field `cells' with the direction of the shock.  The fact that we measure many more sources with $B_{\perp}$ than $B_{\parallel}$ could be evidence of these processes taking place.  On the other hand, $B_{\parallel}$ could be enhanced by shear at the edge of the jet caused by interaction with the surrounding medium (Laing 1981), or by the formation of highly oblique shocks in the jet (Cawthorne \& Cobb 1990).  However, the comparatively low levels of polarisation measured at millimetre/submillimetre wavelengths, suggest that the magnetic field is still predominantly disordered in the core regions of the jets.  This could well be a consequence of the two orthogonal components interacting with each other due to the effects of these `conflicting' physical processes.  In most sources with intermediate levels of polarisation ($3\% < \bar{P} < 9\%$) we have seen magnetic fields gradually changing direction.  This allows us to speculate that there may well be oblique shocks propagating along the jets, or that the apparent direction of the jets could be changing with time, or that the jets are curved on submilliarcsecond scales.  The best way to attempt to sort out the different possible causes of the apparent field rotation is a close examination of the multi-epoch data available on a number of sources.

\subsection{Plane Shocks}
If we start with the basic {\em shock-in-jet} scenario with no bending, then for an initially disordered field, we predict that the formation of a shock results in an increase in the flux level, plus an increase in the polarisation with $B_{\perp}$ being enhanced.  Such behavour is exhibited (to some extent at least) by 0420$-$014, 0735+178, 1921$-$0293, 3C446 and 3C454.3, where the polarisation position angle tends to align itself with the jet when the flux is high, and deviates from alignment as the flux decreases, but only 0735+178 and 3C454.3 show a clear $90\degr$ swing in position angle and a positive correlation between flux and polarisation.  On the other hand, most of the sources are not consistent with this simple picture.  In fact, 0235+164 shows the exact opposite behaviour, where the the magnetic field is perpendicular to the jet when the flux is low and becomes parallel as the flux increases.  In this source, the flux and the polarisation are clearly anti-correlated.  A few other sources (0528+134, 0736+017, 1413+135 and 1749+096) show a similar tendency although the data available is scarce.  A different pattern is observed in 3C279, BL Lacertae and 1335$-$127, whose magnetic fields are {\em always} perpendicular to the jet, despite the fact that all three have undergone very significant changes in flux and polarisation during the period covered by the observations.  What is more puzzling, however, is that the relationship between $P$ and $F$ differs for each source: they are highly correlated in 3C279, anti-correlated in BL Lacertae and in 1335$-$127 the correlation seems to change sign.

If we modify the initial conditions so that the magentic field ordering prior to the formation of the shock is parallel to the jet (possibly due to shearing at the edge of the jet caused by interaction with gas in the host galaxy as suggested by Laing 1993), then we would predict an increase in the flux and a decrease in the polarisation as the shock-enhanced component of the magentic field $B_{\perp}$ cancelled out the polarised emission from the initially parallel field.  The polarisation may then remain very low or could actually rise again, but this time with a position angle orthogonal to the initial direction.  Thus, the overall behaviour could be complex with an initial anti-correlation between flux and polarisation followed by a period where they are correlated after a position angle swing of $90\degr$.

One of the best studied objects in our sample is 3C273 and Stevens {\em et al.} (1997) combine some of our results with centimetre monitoring and VLBI to make a detailed comparison with the predictions of the shock model of Marscher \& Gear (1985).  The most significant observation about 3C273 is that in late 1995 the flux and polarisation both increased as the source flared, however the initially parallel magnetic field {\em remained} parallel.  Examination of the spectral index shows this cannot be due to an optical depth effect.  This behaviour is not therefore consistent with the plane shock model predictions unless the source fortuitously undergoes a very large bend between the scales observed by centimetre VLBI and that of the millimetre/submillimetre emission.

\subsection{Oblique and Conical Shocks}
Lind \& Blandford (1985) suggested that oblique and conical shocks could easily form in relativistic jets and showed that these geometries could have a dramatic effect on predicted number counts.  These authors made little attempt to predict the direct effect of such shock structures on observations of individual objects.  Cawthorne \& Cobb (1990) however, in a prescient but largely-ignored paper, calculated the predicted effects of oblique and conical shock geometries on VLBI imaging and total polarised flux measurements of compact jet sources.  They show that although in general such structures will produce significant polarisation with position angle parellel to the jet (i.e. an enhancement of the perpendicular component of the magnetic field) they can also give rise to the opposite orientation for certain cone opening and viewing angles.  A clear prediction of the model is that when this happens the polarisation will be low ($< 10\%$).

We believe that this type of shock geometry is entirely consistent with the behaviour we observe in many sources: generally low polarisation, often parallel inferred magnetic field and, furthermore, rapid apparent changes between orthogonal magnetic field configurations which can be correlated or anti-correlated with increasing flux and/or polarisation level.  Examination of Figures 2 and 3 of Cawthorne \& Cobb shows that small changes in either cone opening angle, or angle of the jet axis to our line-of-sight, can produce a large swing in apparent magnetic field orientation.  For example, in the case of BL Lacertae $P$ and $F$ are anti-correlated while the position angle changes only slightly about the jet's direction.  OJ287 (another BL Lac object) exhibits the greatest variability in flux and polarisation in the whole sample.  When the polarisation is high its position angle remains fairly unchanged at $\sim 45\degr$ to the jet, despite large changes in flux.  However, on three other occasions it flips from being parallel to perpendicular to the jet.

\subsection{Bending}
There are still sources in which it may be necessary to invoke bending of the jet.  The best example is 3C345, whose jet is known to be very curved even in the highest-frequency VLBI maps.  Our data covers the ocurrence of an outburst and its subsequent gradual decline to the pre-flare level.  The degree of polarisation and position angle, however, show a lot more variability (see Figure~3).  This, together with the fact that the polarisation and the total intensity appear to be largely anti-correlated (see Figure~5), is consistent with a jet being viewed at an initially small angle, which curves away from the line of sight as the shock evolves.  This scenario is supported by the prediction of Gopal-Krishna \& Wiita (1992) that the percentage polarisation and total intensity are anti-correlated as the viewing angle increases within the range determined by the critical value $\theta_{c} = sin^{-1}(1/\Gamma)$, where $\Gamma$ is the Lorentz factor.  In fact, the jet of 3C345 has been estimated to lie within $2\degr$ from our line of sight by Wardle {\em et al.} (1994) and as close as $\sim1\degr$ by Lepp\"{a}nen {\em et al.} (1995) (using the same analysis on a more recent VLBI map).  The relatively low levels of polarisation measured by us on this source ($< 4\%$ on average) are also consistent with a small viewing angle.

It is just possible that a conspiracy of bending on small scales might also reproduce the effects attributed to conical shocks in the previous section for OJ287, for example.  This might then explain the earlier tentative result that $P$ and $F$ tend to be correlated in most quasars and anti-correlated in the BL Lacs as being due to the two groups tending to be viewed either side of the critical viewing angle $\theta_{c}$.  However, it is hard to explain the large degree of flux variability with small or no change in position angle, also commonly seen, in terms of bending. 

We suspect it is far more likely that whilst one or two sources do exhibit significant bending, in most cases the jet merely `wiggles' a little about the axis, producing a modulation of the viewing angle of conical shock structures, whose opening angle also presumably evolves as it travels down an expanding jet.  More theoretical work on the expected evolution of conical shock structures is probably required to make further more detailed comparisons of our results with the ideas of Cawthorne \& Cobb.  Stevens {\em et al.} (1997) have reached a very similar conclusion for the specific example of 3C273.

\section{Conclusions}

\begin{enumerate}
\item The magnetic field on sub-parsec scales in compact jet sources is less highly-ordered than on parsec scales, with 1.1 mm polarisation levels less than 10\% except in a very few sources, where we find it can be up to 15\%.

\item The magnetic field in the most highly-ordered sources is predominantly perpendicular to the jet axis, as measured on parsec scales.

\item In the less-ordered sources, all possible apparent magnetic field orientations have been observed, as well as a high degree of variability of the measured polarisation position angle in the multiply-sampled objects.  These variations are intrinsic and cannot be attributed to optical depth effects.

\item In the best-sampled sources we find a good correlation or anti-correlation between total intensity and degree of polarisation.  In less frequently sampled cases we also generally find indication of such a correlation, and only in a few cases we find no correlation at all, although this may be due to insufficient sampling.

\item Our sample includes similar numbers of BL Lacs and high-polarisation quasars, as well as a few low-polarisation quasars and radio galaxies.  We find, statistically, no difference between the degree of ordering or magnetic field geometry of the BL Lacs and high-polarisation quasars.

\item Although the behaviour of a few sources (again mainly the highly-ordered cases) is very consistent with the predictions of a plane 
{\em shock-in-jet} model such as that of Marscher \& Gear (1985), in most cases the behaviour is inconsistent with this picture, unless very large bends occur on scales smaller than those so far sampled by VLBI. 

\item The behaviour of these sources can be explained by the conical shock model of Cawthorne \& Cobb (1990).  However it may still be necessary to invoke some degree of bending in certain cases.
\end{enumerate}

We have shown the power of a new technique, that of millimetre/submillimetre polarimetry, in deducing new constraints on models of compact, flat-spectrum, radio-jet sources, which have been well-studied using other techniques for many years.  One of the biggest limitations on this experiment has been the uncertainty in the jet direction on the very small (milliparsec) scales sampled at these wavelengths.  Polarisation-sensitive VLBI measurements at 1 mm would be the ultimate tool for this experiemnt, however that is likely to be some years away.  A combination of the highest possible frequency VLBI and millimetre--submillimetre photopolarimetric monitoring with further theoretical development is likely to be the way forward in the near future.

\section*{Acknowledgements}
We thank Jason Stevens of the Joint Astronomy Centre, Hawaii, for useful discussions and for carrying out the 1995 service observing runs.  We also thank Alan Marscher for providing unpublished VLBI information on a number of sources and the anonymous referee for his comments and suggestions that have helped to improve the paper.  The JCMT is operated by the Joint Astronomy Centre on behalf of the UK Particle Physics and Astronomy Council (PPARC), the National Research Council of Canada, the Netherlands Organisation for the Advancement of Pure Research (ZWO) and the University of Hawaii.  RN acknowledges the receipt of an FPI scholarship from the Ministerio de Educaci\'{o}n y Ciencia of Spain.

\bsp

\appendix

\section{Source Summary}

\paragraph*{\em 0133+476 (OC457) \\}
A single observation is available on this source giving a flux density of 1.5 Jy and a percentage polarisation of $10.5\%$.  This is a $6\sigma$ measurement obtained in only 25 minutes integration time on a relatively faint source.  For a $\theta_{VLBI} = 165\degr$ obtained from a low resolution 5 GHz VLBI map, the inferred magnetic field orientation is $18\degr$ away from the perpendicular to the jet.

\paragraph*{\em 0235+164 (OD160) \\}
There are data on this source for four epochs obtained in just over a year.  The flux density changes from about 1 to more than 3 Jy while the polarisation drops from 6 to less than 3\%, indicating a possible anti-correlation between these two parameters ($r=-0.6$ at 60\% significance level).  The mean flux and polarisation over the four epochs are 2.2 Jy and 4.9\% respectively.  A three-point spectral index has been obtained on three epochs giving a shallower than average spectrum (relative to the other sources in the sample).  The position angle flips by about $90\degr$ at the epoch of highest flux.  This is a very compact source with no clear milliarcsecond structure, but for a tentative jet orientation $\theta_{VLBI} = -150\degr$ obtained at 22 GHz, the magnetic field would appear to be perpendicular to the jet during the first two epochs, becoming aligned with it after the flare.

\paragraph*{\em 0316+413 (3C84) \\}
This source is a rather peculiar radio loud seyfert.  Although it is fairly bright (of the order of 2 Jy at 1.1 mm) it seems to have very low polarisation in comparison with most of the sources in the present sample ($\sim 1.5\%$).  This makes it very difficult to achieve polarimetric observations with good signal to noise.  Only two detections have been obtained ten months apart to just over the $2\sigma$ level, which do not show any significant changes in the emission during that period.  High-frequency VLBI maps are available which indicate an inner jet position angle of $-150\degr$; the inferred magnetic field orientation is neither perpendicular nor parallel to the jet.

\paragraph*{\em 0415+379 (3C111) \\}
Although a $\sim 10\sigma$ detection was obtained on this source at all three photometric wavelengths and the source is fairly bright ($>$2 Jy at 1.1 mm) only a $1\sigma$ upper-limmit was measured in polarimetry at the 0.7\% level. 

\paragraph*{\em 0420$-$014 (OF$-$035) \\}
This source was observed four times.  The mean polarisation and flux density are 5\% and 2.8 Jy.  One year separates epochs 2 and 4, during which the 1.1 mm flux drops steadily to almost a quarter.  This does not reflect greatly on the polarisation level but it rotates the position angle by more than $50\degr$.  The core of this source is largely unresolved but very recent 43 GHz VLBI images suggest a jet position angle of $-150\degr$.  This indicates that the magnetic field tends to align itself with the jet as the flare loses strength.

\paragraph*{\em 0430+052 (3C120) \\}
The flux density measured on this radio galaxy at 1.1 mm is 0.75 Jy and only a $1\sigma$ upper-limmit of 2.2\% was obtained in polarimetry.

\paragraph*{\em 0528+134 (OG147) \\}
There are two epochs of observations on this source, taken four months apart, during which the polarisation drops by 40 per cent as the flux almost doubles (increasing from 1.9 to 3.5 Jy) and the position angle suffers a rotation of $80\degr$.  VLBI observations at 22 GHz suggest a jet position angle close to $70\degr$ which indicates that the magnetic field turns towards alignement with the jet as the flux increases.

\paragraph*{\em 0735+178 (OI158) \\}
There are four epochs of data for this source.  The mean polarisation and flux density are 6.3\% and 1.2 Jy.  The milliarcsecond jet orientation is reasonably well determined from VLBI data ($\theta_{VLBI} = 73\degr$) and the magnetic field is seen to change from being perpendicular to being parallel to the jet as the flux decreases from $\sim 1.5$ to 0.8 Jy.  The polarisation also drops during the same period.

\paragraph*{\em 0736+017 (OI061) \\}
This is the faintest source on which we have obtained polarimetric detections: with a flux density of $< 0.6$ Jy the polarisation measured was 5.4\% (with signal to noise $>2\sigma$).  An increase in flux to 0.76 Jy four months later is accompanied with an increase in percentage polarisation to 14.2\% (a $7\sigma$ detection).  Between these two epochs the position angle is rotated by $43\degr$.  The jet position angle observed on a recent 22 GHz map ($\theta_{VLBI} = -80\degr$) suggests that as the flux and polarisation increase, the magnetic field tends towards alignement with the jet. 

\paragraph*{\em 0829+046 \\}
This source has been measured twice within four months, during which the flux density increased from 0.93 to 1.44 Jy and the percentage polarisation decreased from 4.9 to 2.9\%, i.e. they seem to be anti-correlated.  The position angle remained largely the same but the orientation of the magnetic field cannot be infered since no VLBI observations of this source have been found in the literature.

\paragraph*{\em 0851+202 (OJ287) \\} 
This source was observed on eight different occasions and in terms of its polarisation properties it is the most variable source in the sample.  The polarisation ranges widely between 5.1 and $>15\%$ averaging to 9.8\% over all epochs.  However, its measured flux density only varied between 1 and 2.4 Jy.  The milliarcsecond jet position angle has been determined from very-high-frequency VLBI maps to be $\theta_{VLBI} = -120\degr$.  Over most epochs, the magnetic field geometry inferred from the measurements of the polarisation position angle is neither parallel nor perpendicular to the jet.  During the last two epochs, however, the polarisation drops from 13\% to 5\% while the flux increases from 1.3 to 2.4 Jy.  The position angle is clearly seen to rotate by roughly $90\degr$, suggesting that the magnetic field changes from being perpendicular to the jet (at the high polarisation epoch) to being parallel to it.  Over all epochs no correlation is found between the changes in flux and degree of polarisation, but if the data are examined between epochs 1$-$5 and 5$-$8 separately, there is a hint of a positive correlation first $r=0.7$ and an anti-correlation later $r=-0.8$ both with the same 80\% confidence level.

\paragraph*{\em 0923+392 (4C39.25) \\}
There are four polarisation and five photometric epochs on this source.  The variation in polarisation (and flux) is not as dramatic as seen in other sources, ranging between 3.7 and 5.6\% with measured position angles spread over a $\sim45\degr$ range.  However, a flare was recorded between the last two epochs (taken one month apart) when the flux rose from a minimum of 1.1 Jy to a maximum of 2.5 Jy.  The core's position is uncertain since it remains unresolved with VLBI.  Relative to a position of $-88\degr$ for the innermost resolved component, the magnetic field is neither perpendicular nor parallel to the jet, except just before the flare where it is found within $15\degr$ of the aligned orientation.

\paragraph*{\em 1055+018 (4C01.28) \\}
Two observations are available on this source, taken almost three years apart, which are very similar: the polarisation is very high (9\% on average) and the mean flux is $\sim 1.5$ Jy.  The magnetic field appears to be exactly perpendicular to the jet, however the jet position angle ($305\degr$) was obtained from low-frequency maps (1.7 GHz) and it may not represent the true orientation of the milliarcsecond jet.

\paragraph*{\em 1226+023 (3C273) \\}
This source was observed at ten different epochs.  The flux density at 1.1 mm varied in the range 11 -- 21.2 Jy.  This unusually high brightness allows polarisation levels as low as 0.7\% to be measured with enough signal to noise, so far not achieved on any other source.  In the last two epochs (in late 1995) this source was caught flaring and the polarisation rose to 6.8\%.  Except at the epoch of lowest polarisation (and flux) the position angle lies within a small range of $30\degr$.  From numerous high-frequency VLBI observations, we estimate the orientation of the milliarsecond jet to be $\sim -110\degr$, however at 100 GHz the inner jet exhibits a somewhat wiggly morphology.  The inferred magnetic field geometry tends towards alignment with the jet, except at the low polarisation epoch when it is closer to the perpendicular configuration (although with very low level of ordering).  There is also a significant correlation between flux and degree of polarisation with $r=0.8$ at the 99.4\% significance level.

\paragraph*{\em 1253$-$055 (3C279) \\}
This is the most frequently observed source in the sample with 11 polarisation epochs.  It is very bright (with flux density in the range 6 -- 13.5 Jy) and very highly polarised (with an average of 11.3\%).  High-frequency VLBI maps reveal an inner jet position angle of $-135\degr$ within 0.1{\em mas}.  The position angle varies only by a few degrees between epochs, indicating a highly-ordered magnetic field unambiguously perpendicular to the jet.  The flux and polarisation are highly correlated with $r=0.76$ at the 99.4\% significance level.

\paragraph*{\em 1308+326 \\}
A $4\sigma$ observation was obtained on this source giving a percentage polarisation of $\sim 10\%$ and a flux density of 1 Jy.  Using a VLBI jet position angle of $-46\degr$ (obtained at 5 GHz) the inferred magnetic field orientation is neither parallel nor perpendicular.  However, the jet curves significantly within the inner 4{\em mas} and it is likely to bend further within the unresolved region, casting some doubt on the validity of the $\theta_{VLBI}$ used.

\paragraph*{\em 1335$-$127 (1334$-$127) \\}
There are five observations on this source with a flux density in the range 2.5 -- 4.9 Jy, which overall appears to be anti-correlated with the polarisation that ranges between 2.5 and 7.9\%.  The position angle is very steady remaining within a $23\degr$ range over all epochs.
A very tentative jet orientation of $135\degr$ has been obtained from a 5 GHz VLBI map in which the core is not clearly resolved.  This would indicate a magnetic field geometry perpendicular to jet.

\paragraph*{\em 1413+135 \\}
This source has been observed three times, the last two epochs taken within a month.  The flux density ranges between 1.1 and 1.6 Jy and appears to be anti-correlated with the polarisation that varies in the range 3.7 -- 9.1\%.  The only VLBI map available is of low resolution but shows a double-sided jet, the most compact side of which is oriented along $-100\degr$ at $\sim8{\em mas}$.  This is probably a poor estimate
but the indication is that the magnetic field geometry lies roughly parallel to the jet when the polarisation is high.

\paragraph*{\em 1514$-$241 (AP Lib) \\}
This is a low declination source not frequently observed.  The only observation obtained gives a flux density of 1.1 Jy and polarisation of 5.5\% to better than $3\sigma$.  The position angle of the jet has only been estimated from a VLA map with insufficient resolution close to the core to try to establish the orientation of the magnetic field.  This measurement gives a very flat spectral index, in broad agreement with the shallower than average spectral index values obtained on this source by Gear {\em et al.} (1994) over three different epochs.

\paragraph*{\em 1633+382 (4C38.41) \\}
The single observation available on this source was obtained with a flux density $<$0.9 Jy, however the polarisation was high at $\sim 8\%$ with a signal to noise of $3\sigma$.  The polarisation position angle indicates a possibly perpendicular magnetic field relative to a jet orientation of $-45\degr$ determined from a low-frequency (5 GHz) VLBI map.

\paragraph*{\em 1641+399 (3C345) \\}
This source has been observed eight times.  The degree of polarisation changes readily between the extremes of 0.7 and 11.2\%, averaging to 4.7\% over all epochs.  The flux variability is very different, since a large flare at the beginning of the observations (when the flux increases from 1.6 to 5.8 Jy) is followed by a slow decay to its pre-flare level over the subsequent three years.  The jet is known to be very twisted and it is not possible to determine its orientation close to the core with any certainty.  High-frequency VLBI images show that emerging components follow paths in the range $-135$ -- $-60\degr$.  The polarisation position angle changes by tens of degrees between epochs.  Using a recent value for $\theta_{VLBI} = -55\degr$ (obtained at very high frequency) the magnetic field is found within $30\degr$ of alignment with the jet on six epochs, but clearly perpendicular to it on one occasion.  There is a $90\degr$ difference in position angle between the first and last polarisation epochs, when the polarisation was close to 5\% in both cases but the flux measured was 5.8 and 1.9 Jy respectively.

\paragraph*{\em 1741$-$038 (OT$-$68) \\}
There is a single observation on this source with just under $3\sigma$ signal to noise.  The polarisation is 4.6\% with a position angle of $126\degr$ and the flux density is 1.16 Jy.  The inferred magnetic field orientation is parallel to the jet with $\theta_{VLBI} = -155\degr$ (recently observed with VLBI at 22 GHz).

\paragraph*{\em 1749+096 (OT081) \\}
This source was observed on three occasions over 18 months.  The polarisation rises from 3.7 to 8.5\% while the flux drops from 4.6 to 1.9 Jy, i.e. they are anti-correlated.  The position angle varies within a $50\degr$ range, but the compactness of the core makes the jet position angle $65\degr$ used somewhat uncertain and the magnetic field geometry difficult to infer.  

\paragraph*{\em 1823+568 (4C56.27) \\}
There is only one epoch on this source which is highly polarised at 12.8\% and with a flux density of 1.1 Jy.  The magnetic field seems to lie within $25\degr$ from the perpendicular to a jet orientation of $-166\degr$ determined from 5 GHz maps. 

\paragraph*{\em 1921$-$293 (OV$-$236) \\}
This low declination source was observed on four different epochs.  It is very bright, with a flux density ranging from 4.7 to 7 Jy, and highly polarised with a mean polarisation level of 9.1\%.  The position angle changes by tens of degrees between epochs but the magnetic field remains largely perpendicular to the jet (only one VLBI map is available but it has been obtained very recently at 43 GHz).  The flux and degree of polarisation are uncorrelated.

\paragraph*{\em 2200+420 (BL Lacertae) \\}
BL Lacertae has been observed six times.  It is not brighter than the other BL Lacs with a mean flux of 1.8 Jy (varying between 1 and 2.6 Jy) but it is the most highly polarised with a mean polarisation of 8.1\%.  The changes in flux and percentage polarisation are anti-correlated with $r=-0.77$ at 93\% confidence level.  The polarisation position angle is the same on four of the six epochs, thus remaining largely within a level consistent with a magnetic field geometry perpendicular to the jet.  The jet orientation is well determined from several high-frequency VLBI maps.

\paragraph*{\em 2223$-$052 (3C446) \\}
There are two epochs on this source, taken more than two years apart, between which the flux density and the polarisation both drop significantly from 4.5 to 1.1 Jy and from 3.8 to 2.2\% respectively.  The position angle of the jet is $218\degr$ (from a high resolution 100 GHz VLBI map) and suggests a perpendicular magnetic field when the source flux is high, deviating from this position by $50\degr$ when the emission is low.

\paragraph*{\em 2251+158 (3C454.3) \\}
This source has been measured six times.  The flux and polarisation lie in the ranges 3.8 -- 6.7 Jy and 2.9 -- 8.3\% respectively, and are very strongly correlated ($r=0.94$ with a 99.5\%confidence level).  The position angle changes readily within a $>90\degr$ range.  The epoch of lowest flux and minimum polarisation coincide with a magnetic field aligned with the jet, which has a position angle of $-65\degr$ (from 43 GHz VLBI maps).  As the flux and polarisation increase the magnetic field turns towards the perpendicular configuration.

\end{document}